\documentstyle[preprint,epsf,eqsecnum,aps]{revtex}

\begin{document}
\def\appls{\hbox{$<$\kern-.75em\lower 1.00ex\hbox{$\sim$}}}
\draft
\tightenlines
\title{EVIDENCE FOR A NARROW $\sigma(770)$ RESONANCE AND  \\
ITS SUPPRESSION IN $\pi\pi$ SCATTERING FROM MEASUREMENTS OF\\
 $\pi^- p_\uparrow \to \pi^- \pi^+ n$ ON POLARIZED TARGET AT 17.2 GeV/c.}
\author{M. Svec\footnote{electronic address: svec@hep.physics.mcgill.ca}}
\address{Physics Department, Dawson College, Montreal, Quebec, Canada H3Z 1A4\\
and\\
Physics Department, McGill University, Montreal, Quebec, Canada H3A 2T8}
\maketitle
\begin{abstract}
We present a new model independent amplitude analysis of reaction $\pi^- p \to \pi^- \pi^+ n$ measured at CERN at 17.2 GeV/c on polarized target using a Monte Carlo method in an extended range of dipion mass 580--1080 MeV. The two solutions for the measured moduli  $|\overline S|^2\Sigma$ and $|S|^2\Sigma$  of the two $S$-wave transversity amplitudes $\overline S$ and $S$ show resonant behaviour below 800 MeV corresponding to a scalar resonance $\sigma(770)$. Simultaneous fits to $|\overline S|^2\Sigma$ and $|S|^2\Sigma$ including $f_0(980)$ resonance give solution average $m_\sigma = 778 \pm 16$ MeV and $\Gamma_\sigma = 142 \pm 33$ MeV. The CERN data on polarized target supplemented by the assumption of analyticity of production amplitudes in dipion mass allow to determine the helicity amplitudes $S_0$ ($a_1$ exchange) and $S_1$ ($\pi$ exchange) from the fitted transversity amplitudes. The sign ambiguity in $\sigma(770)$ contribution leads to two solutions. In the "down" solution $\sigma(770)$ is suppressed in the helicity flip amplitude $S_1$ and thus also in $\pi^- \pi^+ \to \pi^- \pi^+$ scattering. Most contribution of $\sigma(770)$ to pion production is in the nonflip amplitude $S_0$, or in $\pi^- a_1^+ \to \pi^-\pi^+$ scattering. In the "up" solution the situation is reversed. The "up" solution is excluded by unitarity in $\pi\pi$ scattering. The "down" solution, and thus the evidence for $\sigma(770)$, is in agreement with unitarity in both $\pi^- \pi^+ \to \pi^- \pi^+ $ and $\pi^- a_1^+ \to \pi^- \pi^+$ scattering. There are four "down" solutions $(1, \overline 1)$, $(2, \overline 1)$, $(1, \overline 2)$ and $(2, \overline 2)$. The $\sigma(770)$ resonance manifests itself as a broad resonant structure at $\sim$ 720 MeV in the flip amplitude $|S_1|^2$ in solutions $(1, \overline 1)$ and $(2, \overline 1)$. The contribution of $\sigma(770)$ to $|S_1|^2$ is small in solutions $(1, \overline 2)$ and $(2, \overline 2)$. The metamorphosis of a narrow $\sigma(770)$ in the nonflip amplitude $S_0$ into a broad resonant structure in the flip amplitude $S_1$ is a new phenomenon related to breaking of scale and chiral symmetry in QCD.
\end{abstract}

\section{Introduction}

The existence of a scalar - isoscalar meson $\sigma$ with $\pi^-\pi^+$ decay was first proposed by Schwinger in 1957 in a field theory of strong interactions\cite{schwinger57}.
First experimental evidence for the existence of $\sigma$ meson with a mass near $\rho^0$
came in early 1960's from the measurements of forward - backward asymmetry in $\pi^-\pi^+$
distribution in $\pi^- p \to \pi^- \pi^+ n$ 
production\cite{hagopian63,islam64,patil64,durand65,baton65}. It was expected that the 
$\sigma(750)$ resonance would show up prominently in $\pi^- p \to \pi^0 \pi^0 n$ 
production where the $\rho^0$ does not contribute. However the measurements of this reaction at CERN in 1972 found no clear evidence for $\sigma(750)$\cite{apel72}. Furthermore, in 1973, Pennington and Protopopescu used analyticity and unitarity constraints on partial wave amplitudes in $\pi \pi \to \pi \pi$ scattering (Roy equations) 
to show that a narrow $\sigma(750)$ resonance cannot contribute to $\pi \pi$ 
scattering\cite{pennington73}. From these facts it was concluded that $\sigma(750)$ does not exist and in 1974 Particle Data Group dropped this state from its listings. 

In 1972, van Rossum and his spin physics group at Saclay reported the first measurements of recoil nucleon polarization in $\pi N \to \pi N$ elastic scattering made at CERN at 6 GeV/c\cite{lesquen72}. The resulting complete set of observables enabled the first model independent amplitude analysis of a hadronic reaction\cite{cozzika72}. The results invalidated all Regge models with the exception of the Barger-Phillips $\rho + \rho'$ model. The same group later measured at CERN the polarization in $KN$ charge exchange and determined the tensor $a_2$ exchange amplitudes from this data\cite{fujisaki79}. Not a single Regge model was in agreement with the measured amplitudes. These CERN experiments established that experimental determination of helicity amplitudes in hadron scattering using measurements with spin is crucial for our understanding of hadron dynamics.   
    
In 1978, Lutz and Rybicki extended the concept of amplitude analysis to pion production processes. They showed\cite{lutz78} that almost complete amplitude analysis of reactions $\pi N \to \pi^+ \pi^- N$ and $KN \to K^+ \pi^- N$ is possible from measurements in a single experiment on a transversely polarized target. More recently it was shown\cite{svec97,svec97b} that amplitude analyses of reactions $\pi^- p \to \pi^0 \pi^0 n$ and $\pi^- p \to \eta \pi^- p$ are also possible from measurements on transversely polarized target. The work of Lutz and Rybicki opened a whole new approach to hadron spectroscopy and hadron dynamics by enabling us to study the production of resonances on the level of spin amplitudes rather than spin-averaged cross-sections. The measured spin amplitudes provide more than information about the resonance parameters.  They also connect the resonances to the dynamics of hadron production. The behaviour of resonances on the level of spin amplitudes thus reveals new information about the production mechanism itself. 

The pion production on polarized targets was measured at CERN in $\pi^- p \to\pi^-\pi^+n$
at 17.2 GeV/c\cite{becker79,becker79b,chabaud83,rybicki85,kaminski97},
in $\pi^+ n \to \pi^+ \pi^- p$ at 5.85 and 11.85 GeV/c\cite{lesquen85,svec90,svec92} and in $K^+ n \to K^+ \pi^- p$ also at 5.98 and 11.85 GeV/c\cite{lesquen89,svec89,svec92b}. Recently measurements of $\pi^- p \to\pi^-\pi^+n$ on polarized target at 1.78 GeV/c were made 
at ITEP\cite{alekseev99}. Pion production in $p p \to p \pi^+ n$ using a polarized proton beam was measured at ANL at 6 GeV/c\cite{wicklund86}. 

The CERN measurements of  $\pi^- p \to\pi^-\pi^+n$ and  $\pi^+ n \to \pi^+ \pi^- p$ on polarized targets reopened the question of the existence of $\sigma(750)$ scalar meson. Evidence for a narrow $\sigma(750)$ was found in amplitude analyses of $\pi^- p \to\pi^-\pi^+n$ at 17.2 GeV/c at low momentum transfers $-t=0.005 - 0.20 (GeV/c)^2$ and in  $\pi^+ n \to \pi^+ \pi^- p$ at 5.98 and 11.85 GeV/c at larger momentum transfers
 $-t = 0.2 - 0.4 (GeV/c)^2$\cite{svec92c,svec96,svec97c,svec98}. New evidence for $\sigma(750)$ comes from the amplitude analysis of the ITEP data at 1.78 GeV/c and
$-t = 0.005 - 0.20 (GeV/c)^2$\cite{alekseev99}. The best fit in \cite{svec97c} gives $m_\sigma=753 \pm 19$ MeV and $\Gamma_\sigma = 108 \pm 53$ MeV. These values are in agreement with the ITEP results which give $m_\sigma = 750 \pm 4$ MeV and $\Gamma_\sigma = 119 \pm 13$ MeV.

For dipion masses below 1000 MeV the dipion system is produced in spin states $J=0$ ($S-$wave) and $J=1$ ($P-$wave). There are two $S-$wave and six $P-$wave production amplitudes. The amplitude analysis is carried out in terms of normalized recoil nucleon transversity amplitudes and determines their moduli and cosines of certain relative 
phases\cite{lutz78,svec92}. The two $S-$wave transversity amplitudes $S$ and $\overline S$ can be expressed in terms of nucleon helicity nonflip amplitude $S_0$ and helicity flip amplitude $S_1$. The amplitudes 
$S_0$ and $S_1$ exchange $a_1$ and $\pi$ quantum numbers in the $t-$channel, respectively. 
They are of physical interest since the residue of the pion pole in the flip amplitude $S_1$ is related to the partial wave in $\pi \pi \to \pi \pi$ scattering and the residue of the $a_1$ pole in the nonflip amplitude is related to the partial wave in $\pi a_1 \to \pi \pi$ scattering.  The helicity amplitudes cannot be determined from the measured transversity amplitudes since the CERN data on polarized targets do not give information about the relative phases between the transversity amplitudes. Experimentally, such information can be obtained only in measurements of recoil nucleon polarization.

The measurements of  $\pi^- p \to\pi^-\pi^+n$ on unpolarized target at CERN\cite{grayer74}
were used in model dependent determinations of $\pi\pi$ phase 
shifts\cite{hyams73,esta73,esta74,au87,morgan93}. Various analyses of $\pi\pi$ scattering used these CERN-Munich phase shifts to estimate the mass and the width of the $\sigma$ meson with differing 
results\cite{esta79,kaminski94,tornqvist96,harada96,ishida97,oller97,oller99,igi99,kyoto00}. The CERN-Munich phase shifts were obtained using an enabling assumption that all $a_1$ exchange nonflip amplitudes are zero in pion production. This assumption was invalidated by the CERN measurements of pion production on polarized targets which revealed large and nontrivial $a_1$ exchange amplitudes for any dipion spin\cite{becker79b,svec97c}.

In 1997, Kaminski, Lesniak and Rybicki supplemented the CERN data on polarized target by simplifying assumptions about the absolute phases of $P-$wave amplitudes which enabled them to isolate the helicity flip amplitude $S_1$ and to estimate from it the the $S-$wave phase shift
$\delta^0_0$ in $\pi\pi$ scattering\cite{kaminski97}. They found four solutions for $\delta^0_0$.
The two "steep" solutions showed evidence for a narrow $\sigma(750)$ state.  However the obtained inelasticity $\eta^0_0$ violated unitarity\cite{kaminski97} and contradicted certain 4$\pi$ production data\cite{kaminski00}. The authors concluded that a narrow $\sigma(750)$ is excluded in $\pi\pi$ scattering and thus also in pion production  $\pi^- p \to\pi^-\pi^+n$. Remarkably, their "flat-down" solution was similar to the old CERN-Munich phase shift $\delta^0_0$.

In this work we present a new model independent amplitude analysis of the CERN data on $\pi^- p \to\pi^-\pi^+n$ on polarized target using a Monte Carlo method. The analysis extends the range of dipion mass from 600 - 900 MeV used in the previous study\cite{svec96,svec97c} to 580 - 1080 MeV. The new results for the unnormalized spin "up" amplitude $|\overline S|^2 \Sigma$ show a narrow resonance structure below 880 MeV followed by an enhancement above 900 MeV and a dramatic dip $\sim$ 1000 MeV. The spin "down" amplitude $|S|^2 \Sigma$ shows a broader structure around 700 MeV and a dip at $\sim$1000 MeV. Here  $\Sigma = d^2\sigma/dmdt$ is the integrated cross-section taken from Ref.\cite{grayer74}.

In our next step we supplement the CERN data with an assumption of analyticity of pion production amplitudes in the dipion mass $m^2$. The production amplitudes then satisfy generalized fixed s and t dispersion relations in dipion mass\cite{svec01}. In our finite mass range the analyticity in dipion mass allows us to write the nucleon transversity amplitudes as a sum of Breit-Wigner amplitudes for $\sigma(750)$ and $f_0(980)$ with complex coefficients and a complex coherent background. We use this parametrization to perform simultaneous fits to the measured amplitudes $|\overline S|^2 \Sigma$ and $|S|^2 \Sigma$ to determine the transversity amplitudes $\overline S$ and $S$. We find two different fits, A and B, with the same $\chi^2 /dof$ and the same resonance parameters for $\sigma(750)$. The solution average values are  $m_\sigma = 778 \pm 16$ MeV and $\Gamma_\sigma = 142 \pm 33$ MeV for both fits.

The fixed $s$ and $t$  dispersion relations for transversity production amplitudes determine their absolute phase. We show that the phases of the fitted amplitudes must be the absolute phases of the production amplitudes - up to an overall phase common to all amplitudes. The CERN data supplemented by the analyticity of production amplitudes in dipion mass thus allow a complete determination of the helicity amplitudes from the fitted transversity amplitudes without the measurements of recoil nucleon polarization. This effect of analyticity to reduce the number of spin measurements required for amplitude determination was first observed in an analysis of $\pi N$ charge exchange data. When the data on polarized target for $\pi^- p \to \pi^0 n$ were supplemented by fixed $t$ dispersion relations for the nonflip and flip helicity amplitudes, the fitted helicity amplitudes reproduced the measured helicity amplitudes obtained from measurements of recoil nucleon polarization in $\pi N$ scattering\cite{svec77}. 

There are two solutions for the helicity amplitudes corresponding to the sign ambiguity of the $\sigma(770)$ contribution to the moduli  $|\overline S|^2 \Sigma$ and $|S|^2 \Sigma$. In the "down" solution the contribution of $\sigma(770)$ to the $\pi$ exchange helicity flip amplitude $S_1$ is suppressed. Almost all contribution of $\sigma(770)$ to pion production is in the $a_1$ exchange helicity nonflip amplitude $S_0$. In the "up" solution the situation is reversed with almost all contribution from $\sigma(770)$ being in the $\pi$ exchange amplitude $S_1$. We show that unitarity in $\pi \pi$ scattering clearly excludes this "up" solution. Futhermore, we show that the "down" solution - and thus the evidence for the narrow $\sigma(770)$ resonance - is in full agreement with unitarity in both $\pi^- \pi^+ \to \pi^- \pi^+$ and $\pi^- a_1^+ \to \pi^- \pi^+$ scattering.

We conclude that $\sigma(770)$ is suppressed in $\pi^- \pi^+ \to \pi^- \pi^+$ scattering in agreement with the CERN-Cracow phase shift analysis\cite{kaminski97} and the analyses of $\pi \pi $ scattering using Roy equations\cite{pennington73,colangelo00,anantha01}. The suppression of $\sigma(770)$ in $\pi^- \pi^+ \to \pi^- \pi^+$ scattering however does not imply its absence in pion production processes  $\pi^- p \to\pi^-\pi^+n$ and  $\pi^+ n \to \pi^+ \pi^- p$. Instead, $\sigma(770)$ contributes significantly to pion production due to its strong presence in the $\pi^- a_1^+ \to \pi^- \pi^+$scattering in the nonflip amplitude $S_0$.

There are four "down" solutions corresponding to the four combinations $(i, \overline j)$, 
$i, \overline j =1,2$ of the two solutions for the moduli $|S|^2\Sigma(i)$ and
$|\overline S|^2\Sigma(\overline j)$. In the solution combinations $(1, \overline 1)$ and 
$(2, \overline 1)$ the narrow $\sigma(770)$ manifests itself as a broad resonant structure at 720 MeV in the helicity flip $|S_1|^2$ mass spectrum and thus also in the $S$-wave in $\pi^- \pi^+ \to \pi^- \pi^+$. The metamorphosis of a narrow $\sigma(770)$ in the helicity nonflip amplitude $S_0$ into a broad resonant structure in the helicity flip amplitude $S_1$ in these solutions is a new phenomenon with important connections to symmetries of QCD.   

In a related paper\cite{svec02} we show that Weinberg's mended symmetry\cite{weinberg90} selects solutions $(1, \overline 1)$ and $(2, \overline 1)$. Ellis and Lanik derived a relation between the mass $m_\sigma$ and the partial width $\Gamma(\sigma \to \pi^- \pi^+)$ in an effective field theory with broken scale and chiral symmetry\cite{ellis85}. In Ref.~\cite{svec02} we show that Ellis Lanik relation selects solution $(1, \overline 1)$ and imparts the narrow $\sigma(770)$ resonance a dilaton-gluonium interpretation. The solution $(1, \overline 1)$ gives $m_\sigma = 769 \pm 13$ MeV and $\Gamma_\sigma = 154 \pm 22$ MeV. We thus refer to the $\sigma$ resonance as $\sigma(770)$.

The paper is organized as follows. In Section II we present and discuss the results and reliability of our new amplitude analysis. In Section III we introduce the analyticity of production amplitudes in dipion mass. In Section IV we use it for  the parametrization of unnormalized $S$-wave transversity amplitudes in terms of $\sigma(750)$ and $f_0(980)$ Breit-Wigner amplitudes and coherent backgrounds and present the results of simultaneous fits to amplitudes $|\overline S|^2\Sigma$ and $|S|^2\Sigma$. In Section V we use the fitted transversity amplitudes to determine the helicity amplitudes $S_0$ and $S_1$. In Section VI we show that the"up" solution is excluded by the unitarity in $\pi \pi$ scattering while the "down" solution is in agreement with unitarity in both $\pi \pi$ and $\pi a_1$ scattering. In Section VII we comment on comparisons with $\pi^0 \pi^0$ mass spectra. The paper closes with a summary in Section VIII.
 
\section{Amplitude analysis}

\subsection{Data and results for transversity amplitudes}
 
The high statistics CERN-Munich measurement of $\pi^- p\to\pi^-\pi^+ n$ at 17.2 GeV/c on polarized target was reported in four data sets with kinematics given in the following Table:
 
\narrowtext
\begin{quasitable}
\begin{tabular}{lcccc}
Set & $-t$ & $m$ & $\Delta m$ & Ref.\\
&(GeV/c)$^2$ & (MeV) & (MeV)\\
\tableline
1 & 0.005--0.20 & 600--900 & 20 & [15]\\
2 & 0.01--0.20 & 580--1780 & 40 & [16]\\
3 & 0.005--0.20 & 580--1600 & 20 & [17,19]\\
4 & 0.20--1.00 & 620--1500 & 40 & [18]\\
\end{tabular}
\end{quasitable}

\noindent
In the Table $\Delta m$ is the size of the mass bins. There is only one $t$-bin covering the whole interval of indicated $-t$. 

For invariant masses below 1000 MeV, the dipion system in reactions $\pi N \to \pi^-\pi^+ N$ is produced predominantly in spin states $J=0$ ($S$-wave) and $J=1$ ($P$-wave). The experiments on polarized targets then yield 15 spin-density-matrix (SDM) elements, or equivalently 15 moments, describing the dipion angular distribution\cite{lutz78,svec92}. The measured normalized observables are expressed in terms of two $S$-wave and six $P$-wave normalized nucleon transversity amplitudes. In our normalization
 
\begin{equation}
|S|^2 + |\overline S|^2 + |L|^2 + |\overline L|^2 + |U|^2 + |\overline U|^2 + |N|^2 + |\overline N|^2 = 1
\end{equation}
 
\noindent
where $A=S,L,U.N$ and $\overline A = \overline S, \overline L, \overline U, \overline N$ are the normalized nucleon transversity amplitudes with recoil nucleon transversity ``down'' and ``up'' relative to the scattering plane. The $S$-wave amplitudes are $S$ and $\overline S$. The $P$-wave amplitudes $L$, $\overline L$ have dimeson helicity $\lambda=0$ while the pairs $U$, $\overline U$ and $N$, $\overline N$ are combinations of amplitudes with helicities $\lambda = \pm 1$ and have opposite $t$-channel-exchange naturality. The unnatural exchange amplitudes $L$, $\overline L$, $U$, $\overline U$ receive contributions from ``$\pi$'' and ``$a_1$'' exchanges. The natural exchange amplitudes $N$, $\overline N$ are both dominated by ``$a_2$'' exchange.

Amplitude analysis expresses analytically the eight normalized moduli
 
\begin{equation}
|S|, |\overline S|, |L|, |\overline L|, |U|, |\overline U|, |N|, |\overline N|
\end{equation}
 
\noindent
and six cosines of relative phases
 
\begin{equation}
\cos (\gamma_{SL}), \cos (\gamma_{SU}), \cos(\gamma_{LU})
\end{equation}
\[
\cos (\overline\gamma_{SL}), \cos (\overline\gamma_{SU}), \cos (\overline\gamma_{LU})
\]

\noindent
in terms of the measured observables\cite{lutz78,becker79,svec92}. In (2.3) $\cos\gamma$ and $\cos\overline\gamma$ are cosines of relative phases between pairs of amplitudes with opposite transversity (e.g. between $S$ and $L$, and between $\overline S$ and $\overline L$). The relative phases are not independent

\begin{equation}
\gamma _{SL} - \gamma_{SU} + \gamma_{LU} = 0
\end{equation}
\[
\overline\gamma _{SL} - \overline\gamma_{SU} + \overline\gamma_{LU} = 0
\]

\noindent
and the cosines thus must satisfy relations

\begin{equation}
\cos (\gamma_{SL})^2 + \cos (\gamma_{SU})^2 + \cos(\gamma_{LU})^2 -2\cos (\gamma_{SL}) \cos (\gamma_{SU}) \cos(\gamma_{LU}) = 1
\end{equation}
\[
\cos (\overline\gamma_{SL})^2 + \cos (\overline\gamma_{SU})^2 + \cos(\overline\gamma_{LU})^2 -2\cos (\overline\gamma_{SL}) \cos (\overline\gamma_{SU}) \cos(\overline\gamma_{LU}) = 1
\]

\noindent

There are two similar solutions in each $(m,t)$ bin\cite{lutz78,svec92}. However in some $(m,t)$ bins the solutions are unphysical: either a cosine has magnitude larger than 1 or the two solutions for moduli are complex conjugate with a small imaginary part. Unphysical solutions also complicate error analysis. Two methods are used to find physical solutions for amplitudes and their errors. They are $\chi^2$ minimization method and Monte Carlo method. The two methods are described in\cite{becker79,svec96,svec97b}.

All amplitude analyses in\cite{becker79,becker79b,chabaud83,rybicki85,kaminski97} use $\chi^2$ minimization method to find solution for amplitudes and their errors.
Only Ref.\cite{becker79} reports the results for amplitudes for the Set 1. In\cite{becker79b} and\cite{kaminski97} the ratios $|S|/|\overline S|$ and $S$-wave intensity $I_S$ are given, so it is possible to reconstruct the amplitudes $|S|^2\Sigma$ and $|\overline S|^2\Sigma$ for the Sets 2 and 3. These $S$- wave amplitudes for the Set 2 are shown in Fig.~5 of Ref.\cite{svec97c} and in Fig.~5 of this paper for the Set 3.  

In Ref.\cite{svec96} we used Monte Carlo method for finding the physical solutions for amplitudes and their errors in $\pi^- p \to \pi^- \pi^+ n$ using the data Set 1 for dipion masses in the range 600--900 MeV. In Ref.\cite{svec97c}  we show that the results for the $S$-wave amplitudes obtained using Monte Carlo method agree with the results using $\chi^2$ minimization method in\cite{becker79}.
 
In this work we report results of Monte Carlo amplitude analysis of $\pi^- p \to \pi^- \pi^+ n$ reaction using the data Set 3 in the mass range 580--1080 MeV in order to study the $\sigma(750)-f_0(980)$ interference. Above 1000 MeV our analysis is only approximate as we neglect the $D$-wave. However, below 1080 MeV the $D$-wave contribution is still small and its neglect does not affect the structure of $S$-wave amplitudes. This is confirmed by direct comparison with $S$-wave amplitudes obtained in\cite{kaminski97} using the $\chi^2$ method with $D$-wave included above 980 MeV (see Fig.~3b below). The Monte Carlo method is based on 40,000 selections of spin-density matrix elements within their errors in each $(m,t)$ bin. No physical solution was found in 3 bins -- 650 MeV, 850 MeV and 930 MeV.
 
The amplitude analysis is carried out for normalized amplitudes $|A|^2$ and $|\overline A|^2$, $A = S,L,U,N$. However, from the spectroscopic point of view the relevant information is contained in the unnormalized amplitudes $|A|^2\Sigma$ and $|\overline A|^2\Sigma$ where $\Sigma \equiv d^2\sigma/dmdt$ is the reaction cross-section. It is the unnormalized amplitudes $|A|^2\Sigma$ and $|\overline A|^2\Sigma$ which represent the direct spin-dependent contributions to the mass distribution $d^2\sigma/dmdt$. The unnormalized moduli $|A|^2\Sigma$ and $|\overline A|^2\Sigma$ were calculated using the $\Sigma = d^2\sigma/dmdt$ from Fig.~12 of Ref.\cite{grayer74}.
 
The results of our new Monte Carlo amplitude analysis of $\pi^- p \to \pi^-\pi^+ n$ in the mass range 580--1080 MeV are shown in Fig.~1. First we observe is that the $\rho^0$ peak in $d\sigma/dmdt$ is not uniformly reproduced in all $P$-wave amplitudes. In fact, the amplitudes $|\overline N|^2\Sigma$ and $|U|^2\Sigma$ show considerable suppression of $\rho^0$ production. The shapes of mass distributions with opposite nucleon transversities show considerable differences in both $S$- and $P$-wave amplitudes. The amplitudes $|S|^2\Sigma$, $|L|^2\Sigma$, $|U|^2\Sigma$ with recoil nucleon transversity ``down'' are smaller and broader than the amplitudes $|\overline S|^2\Sigma$, $|\overline L|^2\Sigma$ and $|\overline U|^2\Sigma$ with recoil nucleon transversity ``up''. The opposite is true for the natural exchange amplitudes $|\overline N|^2\Sigma$ and $|N|^2\Sigma$.

The $S$-wave amplitude $|\overline S|^2\Sigma$ shows a resonant behaviour below 880 MeV in both solutions. The relative phase $\overline\gamma_{SL}$ between $\overline S$ and $\overline L$ amplitudes is near zero in Solution 1 and a small constant in Solution 2. Since $|\overline L|^2\Sigma$ clearly resonates, the amplitude $|\overline S|^2\Sigma$ must resonate as well. The amplitudes $|S|^2\Sigma$ and $|L|^2\Sigma$  are both broader and their relative phase $\gamma_{SL}$ is again near zero in Solution 2. This strongly suggests that the amplitude $|S|^2\Sigma$ also resonates.
 
The amplitude $|\overline S|^2 \Sigma$ shows an enhancement between 880 and 980 MeV. Both solutions in both $S$-wave amplitudes dip at 1010 MeV. This dip is accompanied with a dramatic change of phase $\overline\gamma_{SL}$ and $\overline\gamma_{SU}$ above 1000 MeV. This behaviour is interpreted as evidence for a narrow resonance $f_0(980)$. Its contribution must be taken into account in fits to amplitudes $|\overline S|^2\Sigma$ and $|S|^2\Sigma$.

\subsection{$S$-wave intensity} 

The spin-averaged $S$-wave intensity is defined as $I_S = (|S|^2 + |\overline S|^2)\Sigma$. Since there are two independent solutions for the amplitudes $|S(i)|^2$ and $|\overline S(j)|^2$, $i,j=1,2$, we obtain four solutions for $I_S$ which we label as follows

\begin{equation}
I_S (i,j) = (|S(i)|^2 + |\overline S(j)|^2)\Sigma
\end{equation}
 
\noindent
The results are shown in Figure 2. We see that the solutions $I_S(1, \overline1)$ and 
$I_S(2, \overline1)$ are clearly resonating at $\sim$ 770 MeV while the solutions $I_S(1, \overline 2)$ and $I_S(2, \overline 2)$  in addition to a resonant structure show enhancement in the range 880--980 MeV. For comparison we show in Figure 3 the results for $I_S$ in $\pi^- p \to \pi^- \pi^+ n$ at 1.78 GeV/c from the  measurements at ITEP also at $-t$ = 0.005 - 0.20 $(GeV/c)^2$. The data show a clear resonant behaviour at $\sim$ 750 MeV and some energy dependence of the $S$-wave intensity. The $S$-wave intensities in CERN measurements of $\pi^+ n \to \pi^+ \pi^- p$ at 5.98 and 11.85 GeV/c at larger momentum transfers $-t =$ 0.3-0.4 $(GeV/c)^2$ show similarly clear resonant behaviour at $\sim$ 750 MeV\cite{svec96,svec97c}.

\subsection{Reliability of the amplitude analysis}
 
In Monte Carlo amplitude analysis each physical solution obtained from a Monte Carlo selection of input spin density matrix elements satisfies the normalization condition (2.1) and the phase conditions (2.5). The physical solutions are collected and each modulus and cosine has its own range and distribution of values. The mean values of all moduli and cosines of relative phases are calculated independently from the collected physical solutions as the average value of the distribution. If these averages are to represent the measured amplitudes they also must satisfy the conditions (2.1) and (2.5). 

To test the moduli for the normalization condition (2.1) we calculated the l.h.s. of Eq.~(2.1) for the four combinations of solutions $(1,\overline1)$, $(2,\overline1)$, $(1,\overline2)$, $(2,\overline2)$. To test the cosines for the phase conditions (2.5) we calculated the l.h.s. of Eqs.~(2.5) for both solutions $\overline 1$ and $\overline 2$ for $\cos(\overline \gamma)$ and for both solutions $1 $ and $2$ for  $\cos(\gamma)$. The results are shown in Fig.~4. 

Below 850 - 870 MeV the mean values of the moduli satisfy the normalization (2.1) exactly in all four combinations of solutions. Above 850 MeV there are some small deviations except for the combination $(1,\overline1)$ which is still exact. Similarly, the mean values of the cosines satisfy the phase conditions exactly below 870 Mev for all solutions. Above 870 MeV there are small deviations for solution $\overline 2$ of $\cos(\overline \gamma)$ and for solution $1$ for $\cos(\gamma)$ (except one point at 1010 MeV). The deviations for the other solutions are largest around 1000 MeV but are still small. 

The observed deviations are consistent with our neglect of the $D$-wave in our analysis. From Fig.~4 we see that the moduli are less sensitive to the neglect of the $D$-wave than the $S$- and $P$- wave interference terms (the cosines). The agreement of the moduli and cosines with the conditions (2.1) and (2.5) is a powerful test of the self-consistency of the data and the reliability of our amplitude analysis, in particular below 900 MeV where the agreement is exact. 

In Figures 5 and 6 we summarize the results for the $S$-wave amplitudes. Figures~5 and ~6 show the results using the $\chi^2$ minimization  and the Monte Carlo methods of amplitude analysis for the data Set 3, respectively. The results of the two methods for the data Set 1 are shown in Figures 1 and 2 in Ref.~\cite{svec97c}. The two methods are clearly consistent. The data Sets 1 and 3 yield similar results below 900 MeV, both providing model and solution independent evidence for a resonant strucure near 750 MeV. Comparing Figures 5 and 6 for 
the Set 3 we note that the values of $|\overline S|^2 \Sigma$ below 700 MeV are lower in the analysis using $\chi^2$ minimization method. Above 960 MeV the analysis using $\chi^2$ method includes $D$-wave contributions and the results are similar to Monte Carlo analysis which neglects the $D$-wave below 1080 MeV.

\subsection{Interpretation of the measured amplitudes}

We now comment on the interpretation of the measured moduli squared which will be important for their parametrization. To this end we start with the data analysis.
 
The reconstructed and accepted events of $\pi^- p \to \pi^- \pi^+ n$ are grouped in $t$- and $m$- bins. In a given $t$- and $m$-bin each event has its own value of $t$ and $m$. The average values of $t$ and $m$ in a given $t$- and $m$-bin are $t^*$, $m^*$.
 
The dipion angular distribution is described in terms of unpolarized and polarized spin density matrix (SDM) elements. For invariant dipion masses below 1000 MeV the experiments on polarization targets determine 15 SDM elements $\rho_i, i=1, \ldots , 15$ (6 unpolarized, 9 polarized). The SDM elements $\rho_i = \rho_i (s,t, m)$ depend on energy $s$, momentum transfer $t$ and invariant mass $m$. The maximum likelihood optimization in data analysis of events in a given $(t, m)$ bin treats all events as if they had the same $t$ and $m$. The produced SDM $\rho_i$ are assigned to the bin average values $t^*$, $m^*$. Hence the result of the data analysis are values of SDM elements $\rho_i (t^*, m^*)$ at the particular values of $t$ and $m$, namely $t^*$ and $m^*$. The fact that each event in any given $(t,m)$ bin has its own value of $t$ and $m$ is reflected in the errors on $\rho_i$ at $t^*$, $m^*$. 
 
It is a common misconception to think about the measured SDM elements (or, equivalently, moments) as average values of SDM elements over the given $t$- and $m$- bin. We can see that this is not the case by  using the mean value theorem. We get for a bin average of an element $\rho_i$
 
\begin{equation}
{1\over{\Delta t}} {1\over{\Delta m}} \int\limits_{t_1}^{t_2} \int\limits_{m_1}^{m_2} \rho_i (t,m) dtdm = \rho_i (t^*_i, m^*_i)
\end{equation}

\noindent
The bin average of each $\rho_i$ is thus evaluated at its own values of $t^*_i$, $m^*_i$ which are not equal to $t^*$, $m^*$. We then conclude that the measured SDM elements are simply values of $\rho_i$ at the common points $t^*$, $m^*$, namely $\rho_i (t^*, m^*)$.
 
The relations between SDM elements and the moduli of amplitudes and cosines of relative phases given in Ref.\cite{lutz78,becker79,svec92} are valid at any value of $s$, $t$ and $m$. In particular, they are valid at the bin averaged values $t^*$, $m^*$. Hence the analytical solution of these equations\cite{lutz78,becker79,svec92}  determines the moduli of amplitudes and cosines of relative phases also at the bin averaged values $t^*$, $m^*$. We will use this conclusion when we discuss the parametrization of $S$-wave production amplitudes in the Section IV.
 
In the CERN analysis the data sets 1-4 described above all have one $t$-bin and a number of mass bins, $k=1,2,\ldots , N_m$. The average values of $t$ and $m$ in the $k$-th $(t, m)$ bin are $t^*_k, m^*_k$. The values of $t^*_k$ are very close and can be replaced by their average value $t^*$.

\section{Analyticity of Production Amplitudes in Dipion Mass $\lowercase{m}^2$}

Our starting point is the well-known\cite{cushing75} dispersion respresentation of a complex function $f(z)$ with simple poles at $z_n, n=1,2,\ldots , N$ in the complex plane $z$, a branch cut along a positive real axis from $\alpha$ to $\infty$ and with asymptotic property $|z|f(z) \to 0$ as $|z|\to\infty$. We shall also assume that the function $f(z)$ is a real function $f(z^*) = f^*(z)$. Using Cauchy's integral theorem and the process of contour deformation, it can be shown\cite{cushing75} that
\begin{equation}
f(z) = \sum\limits^N_{n=1} {R_n\over{z-z_n}} + {1\over\pi} \int\limits^\infty_\alpha {{Im f(x^\prime) dx^\prime}\over{x^\prime - z}}
\end{equation}
\noindent
A remarkable feature of the proof of (3.1) is that it takes place for a fixed value of $z$\cite{cushing75}. As the result, the dispersion relation (3.1) is also valid for moving poles for which $z_n = z_n (z)$. In such a case the residues $R_n$ in (3.1) also depend on $z$, i.e. $R_n = R_n (z)$. Furthermore, the dispersion relation (3.1) is easily generalized to include a left-hand cut and for functions that are not real. In the latter case $Im f(x^\prime)$ in (3.1) is replaced by a discontinuity function along the cut(s).

The dispersion relation (3.1) can also be generalized to complex functions $f(z,t)$ that depenend on a real parameter $t$. If for values of $t$ in some interval $<t_1,t_2>$ the conditions of the theorem are satisfied, then the function $f(z,t)$ will satisfy generalized fixed-$t$ dispersion relations
\begin{equation}
f(z,t) = \sum\limits^N_{n=1} {R_n(z,t)\over{z-z_n(z)}} + {1\over\pi} \int\limits^\infty_\alpha {{Im f(x^\prime,t) dx^\prime}\over{x^\prime - z}} +l.h.cut(z,t)
\end{equation}
\noindent
Let $s=Re z$ and let us assume that the complex poles $z_n(z)$ have a Breit-Wigner form
\begin{equation}
z_n=s_n(s) = m^2_n - i m_n \Gamma (s),\ n=1, \ldots , N
\end{equation}
\noindent
We define Breit-Wigner amplitudes 
\begin{equation}
a_n(s) = {{-m_n \Gamma_n(s)}\over{s-m^2_n + im_n \Gamma_n(s)}}
\end{equation}
\noindent
and redefine the pole residues to obtain fixed-$t$ dispersion relations   
\begin{equation}
f(s,t) = \sum\limits^N_{n=1} R_n(s,t) a_n (s) + I(s,t)
\end{equation}
\noindent
where $R_n(s,t)$ are the redefined pole residues and $I(s,t)$ are the dispersion integrals over the the left- and right- cuts. Any necessary subtractions are included in the part $I(s,t)$.

An example of the parametric dispersion relations (3.5) are dispersion relations for partial wave amplitudes $T^J(s,t)$ in off-shell $\pi\pi \to \pi\pi$ scattering with off-shell mass squared $t$. Such off-shell partial wave amplitudes are part of the description of $\pi$ exchange production amplitudes in $\pi^- p \to \pi^-\pi^+n$.

The angular distribution of the dipion state $\pi^-\pi^+$ in $\pi^- p \to \pi^-\pi^+n$ is described by partial wave production amplitudes $H^J_{\lambda \lambda_n,0\lambda_p}(s,t,m^2)$ defined by angular expansion of the production amplitudes 
$H_{ \lambda_n,0\lambda_p}(s,t,m^2,\theta,\phi)$ where the angles $\theta$ and $\phi$ describe the direction of the $\pi^-$ in the $\pi^- \pi^+$ rest frame\cite{svec97c,svec01}.
Here $J$ and $\lambda$ are the spin and helicity of the dipion state in the $t$-channel; $\lambda_p$ and $\lambda_n$ are the $s$-channel helicities of the proton and the neutron; $s$ is the c.m.s. energy squared, $t$ is the four-momentum transfer squared and $m$ is the dipion mass. Linear combinations of these amplitudes define nucleon helicity and nucleon transversity amplitudes with definite $t$-channel naturality\cite{lutz78,svec97c}. For masses below 1000 MeV only $S$ and $P$ waves contribute and the participating transversity amplitudes are $A$ and $\overline A$ with $A=S,L,U,N$\cite{lutz78,svec97c}. 

The production amplitudes $A(s,t,m^2)$ and $\overline A(s,t,m^2)$ involve kinematic singularities\cite{martin70} in $t$ and $m^2$. At fixed $s$ and $t$ the relevant singularity in $m^2$ comes from the two-body phase space $q=\sqrt{m^2/4 - \mu^2}$ where $\mu$ is pion mass. Let $F(s,t,m^2)$ and $\overline F(s,t,m^2)$ be the corresponding production amplitudes free from kinematical singularities in $m^2$. Assuming analyticity of the amplitudes $F$ and $\overline F$ in the dipion mass for experimentally accessible values of $s$ and $t$,  we can write fixed $s$ and $t$ dispersion relations
\begin{equation}
F(s,t,m^2) = \sum\limits^N_{n=1} R_n(s,t,m^2) a_n (s) + I(s,t,m^2)
\end{equation}
\noindent
 where $N$ is the number of contributing resonances and $I(s,t,m^2)$ are the dispersion integrals. In a finite mass interval $4\mu^2 \le m^2 \le m^2_M$ where only $M$ resonances contribute we can write\cite{svec01}
\begin{equation}
F(s,t,m^2) = \sum\limits^M_{n=1} R_n(s,t,m^2) a_n (s) + B^{(M)}(s,t,m^2)
\end{equation}
\noindent
 where the complex backcground $B^{(M)}$ 
\begin{equation}
B^{(M)}= \sum\limits^N_{m=M+1} R_m(s,t,m^2) a_m(s) + I(s,t,m^2)
\end{equation}
\noindent
The parametrization (3.7) of production amplitudes in terms of a sum of Breit-Wigner amplitudes with complex coefficients and a complex coherent background has been widely used in fits to production data\cite{alde97,alde98,barberis99,bellazzimi99}. Description of interfering resonances in $\pi\pi$ scattering and in pion production $\pi^- p \to \pi^-\pi^+ n$ and the related questions of unitarity constraints are further discussed in Ref.\cite{svec01}.

Fixed $s$ and $t$ dispersion relations in dimeson mass for production amplitudes were first introduced in 1973 by Hoyer and Kwiecinski\cite{hoyer73} to relate production amplitudes in $a+b \to c+d+e$ to particle-Reggeon scattering $a+R \to c+d$. We comment on the relationship of analyticity in dipion mass and pion-Reggeon scattering in $\pi^- p \to \pi^-\pi^+n$ in Section VI.

\section{Determination of  $S$-wave Transversity Amplitudes from Simultaneous Fits to  $|\overline S|^2\Sigma$ and $|S|^2\Sigma$}
 
\subsection{Parametrization of $S$-wave production amplitudes}
 
To understand the resonant structure of the $S$-wave amplitudes $|\overline S|^2\Sigma$ and $|S|^2\Sigma$ we performed simultaneous fits to these amplitudes using analyticity based  parametrization (3.8) which included a $\sigma(750)$ Breit-Wigner pole, $f_0(980)$ Breit-Wigner pole and a coherent background. Following Ref.\cite{svec97c}, we write the unnormalized $S$-wave amplitudes in the form
 
\begin{equation}
|\overline S|^2\Sigma = q |\overline F|^2
\end{equation}

\[
|S|^2\Sigma = q |F|^2
\]
 
\noindent
where $q$ is the mass dependent phase space factor determined in\cite{svec97c} to be simply the c.m.s. momentum of $\pi^-$ in the $\pi^-\pi^+$ rest frame. The amplitudes $\overline F$ and $F$ are unnormalized amplitudes which are parametrized in terms of resonance contributions and coherent background. For amplitude $F = F(s,t,m)$ we write
 
\begin{equation}
F = R_\sigma (s,t,m) a_\sigma(m) + R_f (s,t,m) a_f (m) + Q (s,t,m)
\end{equation}
 
\noindent
where $a_R$ is the Breit-Wigner amplitude
 
\begin{equation}
a_R = {{m_R \Gamma}\over{m_R^2 - m^2 - im_R \Gamma}}
\end{equation}
 
\noindent
where $m_R$ is the resonant mass, $R = \sigma, f$. In the following $f$ will refer always to $f_0(980)$ resonance. The mass dependent width $\Gamma(m)$ depends on spin $J$ and has a general form
 
\begin{equation}
\Gamma = \Gamma_R ({q\over q_R})^{2J+1} {{D_J (q_R r)}\over{D_J (qr)}}
\end{equation}
 
\noindent
In (4.4) $q_R = q (m = m_R)$ and $D_J$ are the centrifugal barrier functions of Blatt and Weiskopf\cite{blatt52}
 
\begin{equation}
D_0 (qr) = 1.0
\end{equation}

\[
D_1 (qr) = 1.0 + (qr)^2
\]
 
\noindent
where $r$ is the interaction radius. We recall that
 
\begin{equation}
Re\  a_R = ({{m^2_R - m^2}\over{m_R\Gamma}}) |a_R|^2 \equiv w_R |a_R|^2\ 
\end{equation}
\[
Im\ a_R = |a_R|^2\equiv H_R\
\]

\noindent
In (4.2) the term $Q(s,t,m)$ is the coherent nonresonant complex background and $R_\sigma(s,t,m)$ and $R_f(s,t,m)$ are complex coupling functions. The amplitude $\overline F$ has a form similar to (4.2) with obvious replacements $R_\sigma \to \overline R_\sigma$, $R_f \to \overline R_f$ and $Q\to \overline Q$.
 
The energy variable $s$ is fixed and will be omitted in the following. It is convenient to factor out the phase of $R_\sigma$ and define
 
\begin{equation}
R_\sigma = |R_\sigma|e^{i\phi} = \sqrt {N_S} e^{i\phi}
\end{equation}
 
\[
C=R_f/R_\sigma = C_1 + iC_2
\]

\[
B = Q/R_\sigma = B_1 + iB_2
\]
 
\noindent
Then the unnormalized amplitude (4.2) has the form
 
\begin{equation}
F = \sqrt {N_S} \{ a_\sigma + Ca_f + B\} e^{i\phi}
\end{equation}
 
\noindent
and the parametrization of $|S|^2\Sigma$ reads

\begin{equation}
 |S|^2\Sigma = q N_S\{ [1 + 2w_\sigma B_1 + 2B_2] H_\sigma + B_1^2 + B_2^2 + [C_1^2 + C_2^2] H_f +
\end{equation}
\[
+ 2 [(w_\sigma H_\sigma + B_1) (w_f C_1 - C_2 ) +  (H_\sigma + B_2) (C_1 + w_f C_2)] H_f\}
\]
 
\noindent
The parametrization of $|\overline S|^2\Sigma$ has the same form as (4.9) with the obvious replacements $N_S \to \overline N_S$, $B_i \to \overline B_i$, $C_i \to \overline C_i$, $i=1,2$.
 
The equation (4.9) for $|S|^2\Sigma$ and the analogous equation for $|\overline S|^2\Sigma$ are valid for any $s$, $t$ and any $m \appls$ 1000 MeV. As discussed in the Section II.C, the amplitude analysis produces values of $|S|^2\Sigma$ and $|\overline S|^2\Sigma$ at points ($t^*, m^*_k)$, $k=1,2,\ldots N_m$. In our case $t^* = 0.069$ (GeV/c)$^2$ and $N_m = 24$ bins. Evaluating the r.h.s. of (4.9) and the corresponding equation for $|\overline S|^2\Sigma$ at $(t^*, m^*_k)$ yields (at a given $s$) values

\begin{equation}
N_S (t^*, m^*_k), B_i (t^*, m^*_k), C_i (t^*, m^*_k) 
\end{equation}
\[
\overline N_S (t^*, m^*_k), \overline B_i (t^*, m^*_k), \overline C_i (t^*, m^*_k)
\]

\noindent
with $i=1,2$ and $k=1,\ldots , N_m$. To perform the fit to data on $|S|^2\Sigma$ and $|\overline S|^2\Sigma$ we must make assumptions about the mass dependence of functions $N_S$, $B_i$, $C_i$, $\overline N_S$, $\overline B_i$, $\overline C_i$, $i=1,2$. We will assume that they depend only weakly on the dipion mass $m$ and can be taken as mass independent in our fits. Then
 
\begin{equation}
N_S (t^*), B_i (t^*), C_i (t^*), \overline N_S (t^*), \overline B_i (t^*), \overline C_i (t^*) 
\end{equation}

\noindent
are simply constants. They are the free parameters to be determined by our simultaneous fits to $|S|^2\Sigma$ and $|\overline S|^2\Sigma$ data. Notice that the parameters (4.11) are values of functions $N_S$, $B_i$, $C_i$, $\overline N_S$, $\overline B_i$, $\overline C_i$, $i=1,2$ at $t=t^*$ and not average values of these functions over the $t$-bin as assumed in Ref.\cite{svec97c}.
  
\subsection{Results of simultaneous fits to $|\overline S|^2\Sigma$ and $|S|^2\Sigma$}
 
Hadron resonance in a production process is represented as a Breit-Wigner pole in helicity or transversity amplitudes with the mass and width independent of helicities or transversities of external particles. We assume that the $\sigma$ resonance contributes to both transversity amplitudes $\overline F$ and $F$ and that the difference in the shape of $|\overline S|^2\Sigma$ and $|S|^2\Sigma$ is due to the difference in coherent background and the $\sigma(750)-f_0(980)$ couplings. The measured mass distributions $|\overline S|^2\Sigma$ and $|S|^2\Sigma$ were augmented at 650 and 850 MeV mass bins by data points from Monte Carlo amplitude analysis using the data Set 1\cite{svec96}. 

We used the parametrization (4.9) for $|\overline S|^2\Sigma$ and $|S|^2\Sigma$ with common $m_\sigma$, $\Gamma_\sigma$, $m_f$ and $\Gamma_f$. We used the programme FUMILI\cite{silin85} to perform a simultaneous fit to data on $|\overline S|^2\Sigma$ and $|S|^2\Sigma$. The two solutions for $|\overline S|^2\Sigma$ are labeled $\overline 1$ and $\overline 2$, while the two solutions for $|S|^2\Sigma$ are labeled 1 and 2. Thus there are 4 simultaneous fits $(1, \overline1), (2, \overline 1), (1, \overline2)$ and $(2, \overline 2)$. 

We used 15 different initial values of the free parameters, including fits with a fixed mass and width of $f_0(980)$ at $m_f = 980$ MeV and  $\Gamma_f = 40$ MeV. Improved fits were obtained when $m_f$ and $\Gamma_f$ were allowed to be free parameters. Two solutions, labeled Fit A and Fit B, were found for the free parameters giving the same values of resonance parameters $m_\sigma$, $\Gamma_\sigma$, $m_f$, $\Gamma_f$ and the same values of $\chi^2/dof$ for each combination of solutions for the amplitudes. Most initial values produced the Fit B.  

The results of the best fits are shown in Fig.~6. The two fits A and B produce virtually identical curves. The fitted values of $m_\sigma, \Gamma_\sigma, m_f$ and $\Gamma_f$ and the associated $\chi^2/dof$ for each combination of solution are given in Table I. The results for the normalization parameters $\overline N_S$, $N_S$, the background parameters $\overline B_i$, $B_i$, $i=1,2$ and the $f_0(980)$ coupling parameters $\overline C_i$, $C_i$, $i=1,2$ are given in the Tables II and III for the Fit A and Fit B, respectively. We observe that the free parameters for the amplitude $|S|^2|\Sigma$ are essentally the same for both fits A and B. For the amplitude $|\overline S|^2\Sigma$ the fits A and B differ chiefly in the values of $\overline N_S$, $\overline B_2$ and especially in $\overline C_1$ and $\overline C_2$. 

We see from Fig.~6 that the parametrization (4.9) reproduces well the structure of $S$-wave amplitudes $|\overline S|^2\Sigma$ and $|S|^2\Sigma$ in both solutions. The $\sigma$ resonance peaks are clearly visible in both solutions to $|\overline S|^2\Sigma$ while the presence of $\sigma$ resonance in $|S|^2\Sigma$ is required to explain the broad structure observed in this amplitude. The interference of $f_0(980)$ with coherent background and the $\sigma$ resonance is responsible for the enhancements in the mass region 880--980 MeV, the rapid decrease at 990 MeV and the dip at 1010 MeV in all solutions.
 
We see in Fig.~6 that the cross-fits $(2, \overline 1)$ and $(1, \overline 2)$ are similar to fits $(1, \overline 1)$ and $(2, \overline 2)$, respectively. In Table I we notice that all fits have similar values of $m_\sigma, \Gamma_\sigma, m_f$ and $\Gamma_f$. The average values of mass and width of $\sigma$ resonance are
 
\begin{equation}
m_\sigma = 778 \pm 16\ {\rm MeV}\ ,\ \Gamma_\sigma = 142 \pm 33\ {\rm MeV}
\end{equation}
 
\noindent
This compares with $m_\sigma = 753 \pm 19$ MeV and $\Gamma_\sigma = 108 \pm 53$ MeV obtained in\cite{svec97c} from fits only to $|\overline S|^2\Sigma$ without the interference with $f_0(980)$. The simultaneous fit of $|\overline S|^2\Sigma$ and $|S|^2\Sigma$ and the interference with $f_0(980)$ thus require a slightly higher mass and a broader width of the $\sigma$ state. In the following we refer to the $\sigma$ meson as $\sigma(770)$ resonance.

The average values of mass and width of $f_0(980)$ resonance determined by our fits are
\begin{equation}
m_f = 973 \pm 12\ {\rm MeV}\ ,\ \Gamma_f = 51 \pm 25\ {\rm MeV}
\end{equation}
This compares well with the Particle Data Group estimates of $m_f = 980 \pm 10$ MeV and $\Gamma_f = 40-100$ MeV.
 
The program FUMILI calculates the $\chi^2$ contribution for each data point and  $\chi^2/dof$ for each solution combination $(i, \overline j)$, $i, \overline j = 1,2$. The results are presented in the Table I and are the same for both fits A and B. The values of $\chi^2/dof$ show very good fits to all data with an overall average of $\chi^2/dof = 0.4420$.

\subsection{The consistency of $\sigma(770)$ with relative phases between $S$-wave and $P$-wave amplitudes}

The moduli of $S$-wave amplitudes $|\overline S|^2\Sigma$ and $|S|^2\Sigma$ provide model and solution independent  evidence for a $\sigma$ state below 800 MeV and we have used them to determine its resonance parameters. Additional evidence for the $\sigma$ resonance is provided by the cosines of relative phases between the $S$-wave and $P$-wave amplitudes shown in Figures 1c and 1d.

Below 900 MeV the cosines $\cos (\overline \gamma_{SL})$ and  $\cos (\gamma_{SL})$ are nearly equal to +1 in Solution 1 and are close to +1 in Solution 2. The pairs of amplitudes $(\overline S, \overline L)$ and $( S,L)$ are thus in phase in Solution 1 and nearly in phase in Solution 2. Since the amplitudes $|\overline L|^2\Sigma$ and $|L|^2\Sigma$ resonate near 770 MeV due to the presence of $\rho^0$, a resonance with a mass and width similar to $\rho^0$ is expected in the amplitudes $\overline S$ and $S$.

Below 900 MeV the cosines $\cos (\overline \gamma_{SU})$ and $\cos (\gamma_{SU})$ are both nearly constant and close to -1 in both solutions. The pairs of amplitudes
 $(\overline S, \overline U)$ and $( S,U)$ are thus nearly $180^{\circ}$ out of phase. Although the $\rho^0$ production is suppressed in the amplitudes $|\overline U|^2\Sigma$ and $|U|^2\Sigma$ compared to $|\overline L|^2\Sigma$ and $|L|^2\Sigma$, these amplitudes clearly resonate near 770 MeV and the near constant phase lag near $180^{\circ}$ is again consistent with a $\rho$-like $\sigma$ state in the $S$-wave amplitudes.

If the pairs of amplitudes $(\overline S, \overline L)$ and $(S,L)$ are in phase and the pairs  $(\overline S,\overline U)$ and $(S,U)$ are  nearly $180^{\circ}$ out of phase, then the pairs of $P$-wave amplitudes  $(\overline L, \overline U)$ and $(L,U)$ must also be nearly $180^{\circ}$ out of phase and the their cosines should be nearly constant and close to -1 below 900 MeV. In Figures 1c and 1d we see that this is the case. From Fig.~4 we see that the cosines satisfy the phase conditions (2.5). As discussed in Section II.B , this self-consistency between the relative phases is an important test of the self-consistency of the experimental data and the validity of our model independent amplitude analysis. 
 
It is noteworthy that the relative phases in the amplitude analyses of $\pi^-p \to \pi^-\pi^+n$ at 1.78 GeV/c for $-t=$ 0.005--0.20 $(GeV/c)^2$ \cite{alekseev99} and in $\pi^+n \to \pi^+\pi^- p$ at 5.95 $(GeV/c)^2$ and 11.85 $(GeV/c)^2$ for $-t=$ 0.2 --0.3 $(GeV/c)^2$ show the same self-consistency of phases and also find that the $(\overline S, \overline L)$ and $( S,L)$ are in phase in Solution 1 and nearly in phase in Solution 2.  These independent results strengthen further the evidence for the $\sigma(770)$ resonance.   

This qualitative discussion relating the evidence for the $\sigma$ resonance with the measured relative phases can be translated into a quantitative analysis involving simultaneous fits to the moduli

\begin{equation}
|\overline S|^2\Sigma, |S|^2\Sigma, |\overline L|^2\Sigma, |L|^2\Sigma
\end{equation}

\noindent
and interference terms

\begin{equation}
\overline X_{SL}=|\overline S||\overline L| \Sigma \cos (\overline \gamma_{SL}) = Re (\overline S  \overline L^*) \Sigma 
\end{equation}

\[
X_{SL}=|S||L| \Sigma \cos (\gamma_{SL}) = Re (SL^*) \Sigma 
\]
\noindent
using parametrizations (4.2) for the $S$- wave amplitudes and similar parametrizations with  $\rho^0(770)$ in the amplitudes $\overline L$ and $L$ for dipion masses below 1080 MeV. The results of such an analysis could be used for a new  determination of  the $S$- and $P$- wave phase shifts $\delta^0_0$ and $\delta^1_1$ in $\pi \pi$ scattering.

\section{Separation of  $\pi$ and $\lowercase {a}_1$ exchange helicity amplitudes.}

\subsection{Analyticity and the phases of transversity amplitudes $S$ and $\overline S$}

The $S$- wave transversity amplitudes $S$ and $\overline S$ are a linear combination of nucleon helicity non-flip $a_1$ exchange amplitude $S_0$ and nucleon helicity flip $\pi$ exchange amplitude $S_1$. The inverse relations are\cite{svec97c}

\begin{equation}
S_0 = {1\over{\sqrt 2}} (\overline S + S)
\end{equation}
\[
S_1 = {i\over{\sqrt 2}} (\overline S -S)
\]
\noindent
In this Section the amplitudes $S$, $\overline S$, $S_0$ and $S_1$ are understood to be the unnormalized amplitudes. To determine the helicity amplitudes we need to know the phases of the transversity amplitudes. Experimental determination of the relative phase between $S$ and $\overline S$ requires difficult measurements of recoil nucleon polarization\cite{lutz78}. Here we show that the analyticity of production amplitudes in dipion mass determines fully the phases of transversity amplitudes, up to a sign ambiguity that will be resolved using unitarity in $\pi\pi $ scattering.

Fixed-$t$ dispersion relations for helicity amplitudes in two-body scattering fix their absolute phases. Amplitude analysis of $\pi^- p \to \pi^0 n$ charge exchange using fixed-$t$ dispersion relations determined the helicity helicity non-flip and flip amplitudes $F_0$ and $F_1$ from fits to differential cross-section $d\sigma / dt$ and polarization $P$ over a range of energies\cite{svec77}. We note that

\begin{equation}
{d\sigma \over dt} = |F_0|^2 + |F_1|^2
\end{equation}
\[
P{d\sigma\over dt} = 2Re (F_0F_1^*)
\]
\noindent
The fitted amplitudes reproduce the measured amplitudes obtained using measurements of recoil nucleon polarization. This suggests that analyticity can be used to replace the unknown information about the recoil nucleon polarization.

In $\pi^- p \to \pi^- \pi^+ n$ the partial wave intensity $I_A$ and partial wave polarization $P_A$ are given by equations similar to (5.2)

\begin{equation}
I_A = |A_0|^2 + |A_1|^2 = |A|^2 + |\overline A|^2
\end{equation}
\[
P_AI_A= 2 \epsilon Re (A_0A_1^*) = |A|^2 - |\overline A|^2
\]
\noindent
where $\epsilon = +1$ for $A=S,L,U$ and $\epsilon = -1$ for $A=N$. The fixed $s$ and $t$ dispersion relations for the production amplitudes also fix their absolute phases. The analogy with $\pi N$ scattering suggests that using these dispersion relations to determine production amplitudes from polarized target data will also yield true amplitudes.

The parametrizations (4.2) follow from the fixed $s$ and $t$ dispersion relations (3.7) and thus we expect the phases of amplitudes in (4.2) to be their absolute phases. However the fits are made to the moduli and the fitted transversity amplitudes $S_{fit}$ and $\overline S_{fit}$ have their own definite phases. Let us suppose that the true amplitudes $S$ and $\overline S$ differ in phase

\begin{equation}
S = S_{fit} e^{i \phi}\ , \ \overline S = \overline S_{fit} e^{i \overline \phi}
\end{equation}
\noindent
We note that 

\begin{equation}
Re (S \overline S^*) = Re (S_{fit} \overline S_{fit}^*) \cos\delta - Im (S_{fit} \overline S_{fit}^*) 
\sin\delta
\end{equation}
\[
Re (\overline S S^*) = Re (S_{fit} \overline S_{fit}^*) \cos\delta + Im (S_{fit} \overline S_{fit}^*) 
\sin\delta
\]
\noindent
where $\delta = \overline \phi - \phi$. Since $Re (S \overline S^*)=Re (\overline SS^*)$ we find 
$\sin \delta = 0$ which means that 
\begin{equation}
\overline \phi = \phi 
\end{equation}
\noindent
or
\begin{equation}
\overline \phi = \phi \pm \pi
\end{equation}
\noindent
The amplitudes $S$, $\overline S$, $S_0$ and $S_1$ thus share a common phase $e^{i\phi}$. It is easy to show that this phase is common also to all $P$-wave (and higher) amplitudes $A$ and $\overline A$. The absolute phase $e^{i\phi}$ can thus be omitted and the helicity amplitudes $S_0$ and $S_1$ can be calculated using either the pair $(S_{fit}, + \overline S_{fit})$ corresponding to (5.6) or $(S_{fit}, - \overline S_{fit})$ corresponding to (5.7) for the amplitudes $S$ and $\overline S$ in (5.1). 

\subsection{Helicity amplitudes $S_0$ and $S_1$}

The results for the helicity amplitudes corresponding to the phases (5.6) with $S=S_{fit}$ and $\overline S=+\overline S_{fit}$ are shown in Fig.~7. For each solution combination of amplitudes $|S|^2 \Sigma$ and $|\overline S|^2 \Sigma$ we present the real and imaginary parts of the helicity amplitudes $S_0$ and $S_1$ and their moduli $|S_0|^2$ and $|S_1|^2$ for both fits A and B.

The immediate and most striking observation is that $\sigma(770)$ is present almost entirely in the $a_1$ exchange amplitude $S_0$. The contribution of $\sigma(770)$ to the $\pi$ exchange amplitude $S_1$ is severely suppressed. The $S$-wave pion production in $\pi^- p \to \pi^- \pi^+ n$ below 900 MeV is dominated by the $a_1$ exchange amplitude $S_0$ and $\sigma(770)$.

We now look at amplitudes $S_0$ and $S_1$ calculated using the the pair $(S_{fit}, - \overline S_{fit})$ for the amplitudes $(S, \overline S)$ in (5.1). This corresponds to the phases (5.7). In the following we briefly label $S_0(+)$, $S_1(+)$ the helicity amplitudes obtained from the pair   $(S_{fit}, + \overline S_{fit})$ and  $S_0(-)$, $S_1(-)$ the helicity amplitudes obtained from the pair   $(S_{fit}, - \overline S_{fit})$. From (5.1) we find

\begin{equation}
Re S_0(-) = - Im S_1(+)\ , \ Im S_0(-) = + Re S_1(+) 
\end{equation}
\[
Re S_1(-) = + Im S_0(+)\ , \ Im S_1(-) = - Re S_0(+) 
\] 
\noindent
We see that the roles of $S_0$ and $S_1$ have been interchanged. It is now the helicity flip amplitude $|S_1(-)|^2$ which dominates the pion production and receives most of the $\sigma(770)$ contribution. We thus encounter an analogue of the "up - down " ambiguity known from the CERN-Munich and CERN-Cracow determinations of $\pi\pi$ phase shifts. In our case the "down" solution are the amplitudes $S_n(+)$ and the "up" solution are the amplitudes $S_n(-)$, $n=1,2$ in all four solution combinations $(i, \overline j), i, \overline j =1,2$ of moduli $|S|^2 \Sigma$ and $|\overline S|^2 \Sigma$. The origin of the "up - down" ambiguity in our case is the sign ambiguity of $\overline S = \pm \overline S_{fit}$.

\subsection{Relative contribution of $\sigma(770)$ to helicity amplitudes}

To examine the quantitative contribution of $\sigma(770)$ to helicity amplitudes we now write the transversity amplitudes in the form

\begin{equation}
\overline S = q(\overline R_\sigma a_\sigma + \overline R_f a_f + \overline Q)
\end{equation}
\[ 
S = q(R_\sigma a_\sigma + R_f a_f + Q)
\]

\noindent
where $\overline R_\sigma > 0$, $R_\sigma > 0$ for "down"solution. Here  

\begin{equation}
\overline R_\sigma = \sqrt {\overline N_S} \ , \ \overline R_f=\overline R_\sigma \overline C\ , \ \overline Q = \overline R_\sigma B
\end{equation}
\[
R_\sigma = \sqrt {N_S}\ , \ R_f=R_\sigma C\ , \ Q = R_\sigma B
\]

\noindent
are expressed in terms of fitted parameters. The helicity amplitudes then have a form

\begin{equation}
S_0 = q( R_{\sigma 0}a_\sigma + R_{f 0}a_f +Q_0)
\end{equation}
\[
S_1 = iq( R_{\sigma 1}a_\sigma + R_{f 1}a_f +Q_1)
\]

\noindent
where the $\sigma$ couplings to helicity amplitudes are

\begin{equation}
R_{\sigma 0}={1 \over \sqrt 2}(\overline R_\sigma +R_\sigma)\ , \  R_{\sigma 1}={1 \over \sqrt 2}(\overline R_\sigma - R_\sigma)
\end{equation}
\noindent
and similar equations for $R_{f n}$ and $Q_n$, $n=1,2$. The results for $R_{\sigma 0}$ and $R_{\sigma 1}$ are shown in Table IV. We find that the $\sigma$ couplings to flip amplitude $S_1$ are small. They are largest and negative in the $(2,\overline 1)$ solution combination and smallest and postive in the $(2,\overline 2)$ solution combination.

The smallness of $R_{\sigma 1}$ suggests that $\sigma(770)$ may be absent in the flip amplitude. We tested this assumption in modified fits $A_m$ and $B_m$ under a constraint that $N_S = \overline N_S$, i.e. with $R_{\sigma 1} = 0$. The most notable difference with the fits A and B is the flat behaviour of $|S_1|^2$ below 900 MeV shown in Fig.~7 for the Fit $A_m$. The fits $A_m$ and $B_m$ are close to fits A and B but suffer from somewhat larger values of $\chi^2/dof$. This indicates that some contribution of $\sigma(770)$ to the flip amplitude may still be necessary. 

Figure 7 shows important distinctions between the pairs of solution combinations 
$(1, \overline 1)$, $(2, \overline 1)$ and $(1, \overline 2)$, $(2, \overline 2)$ in the behaviour of $|S_1|^2$ mass spectrum below 900 MeV. In solution combinations $(1, \overline 1)$, 
$(2, \overline 1)$ we observe a broad and low resonant structure centered at $\sim$ 720 MeV in fits A and B clearly above the nonresonating fit $A_m$. In solution combinations $(1, \overline 2)$, $(2, \overline 2)$ the fits A and B show no clear resonant structure and nearly coincide with the nonresonating fit $A_m$. From Table I we find that the average mass of $\sigma$ resonance in the solution combinations $(1, \overline 1)$, $(2, \overline 1)$ is 
$m_\sigma = 772 \pm 12 MeV$.

The broad resonant structure at 720 MeV seen in $|S_1|^2$ in the solution combinations
$(1, \overline 1)$, $(2, \overline 1)$ in both fits A and B indicates that a narrow resonance can contribute to a production amplitude without a marked resonance behaviour such as a prominent peak or a dip. In our case the narrow  $\sigma(770)$ resonance appears as a narrow resonance in the non-flip amplitude $S_0$ but manifests itself as a broad resonant structure in the flip amplitude $S_1$. This metamorphosis of a narrow $\sigma(770)$ in the nonflip amplitude $S_0$ into a broad resonant structure in the flip amplitude $S_1$ is a new phenomenon related to breaking of scale and chiral symmetry in QCD\cite{svec02}.

\subsection{Relative contribution of $f_0(980)$ to helicity amplitudes}

Figure 7 shows also interesting distinctions between the pairs of solution combinations 
$(1, \overline 1)$, $(2, \overline 1)$ and $(1, \overline 2)$, $(2, \overline 2)$ in the behaviour of $|S_0|^2$ and $|S_1|^2$ mass spectra above 900 MeV. In this mass region we also find the differences between the fits A and B.

Looking at the nonflip spectrum $|S_0|^2$, the $f_0(980)$ resonance appears as a shoulder or a small peak in solution combinations $(1, \overline 1)$, $(2, \overline 1)$ and as a pronounced narrow peak in solution combinations $(1, \overline 2)$, $(2, \overline 2)$. The fits A and B are distinguished by the height of the $f_0(980)$ peaks.

In the flip spectrum $|S_1|^2$ the $f_0(980)$ resonance appears as a shallow dip in solution combinations $(1, \overline 1)$, $(2, \overline 1)$ and as a pronounced narrow dip in solution combinations $(1, \overline 2)$, $(2, \overline 2)$ in the Fit A. In the Fit B such dips are preceeded by a peak at $\sim$ 920 - 950 MeV which is more pronounced in the solution combinations $(1, \overline 2)$ and $(2, \overline 2)$.

\section{Consistency of $\sigma(770)$ with unitarity in $\pi^- \pi^+ \to \pi^- \pi^+$ and $\pi^- \lowercase{a}_1^+ \to \pi^- \pi^+$ scattering} 

In this Section we show that the "up" solution $S_n(-), n=1,2$ is excluded by the unitarity in $\pi\pi$ scattering while the "down" solution $S_n(+)$ is allowed.  Consequently $\sigma(770)$ is suppressed in $\pi^- \pi^+ \to \pi^- \pi^+$ reaction.

Consider a nucleon $s$-channel helicity flip amplitude in $\pi^- p \to \pi^-\pi^+n$ with a dipion spin J and $t$-channel helicity $\lambda = 0$\cite{svec97c}

\begin{equation}
A^J_1(s,t,m^2) = H^J_{0+,0-}(s,t,m^2)
\end{equation}
\noindent
The pion exchange amplitudes $A^J_1$ have a general form

\begin{equation}
A^J_1 = {R^J_1(s,t,m^2)\over {t-\mu^2}} + M^J_1(s,t,m^2)
\end{equation}
\noindent
where $R^J_1$ is the pion pole residue and $M^J_1$ is the non-pole term. The connection to 
$\pi\pi$ scattering is provided by a factorization hypothesis

\begin{equation}
R^J_1 = - F^J_1(s,t)T^J_\pi(m^2,t,0) 
\end{equation}
\noindent
where $T^J(m^2,t,0)$ is the off-shell partial wave amplitude in $\pi^- \pi^+ \to \pi^- \pi^+$ scattering with the exchanged pion having mass t and spin 0. The factor $F^J_1$ is positive. It contains the $\pi N N$ form factor, phase space factors\cite{svec97c} and angular momentum constraints. We have verified explicitely that the factorization condition (6.3) is consistent with unitarity conditions for partial wave production amplitudes in $\pi^- p \to \pi^- \pi^+ n$ derived in\cite{svec01}. The consistency requires that the helicity nonflip amplitude $H_{0+,0+}$ in $\pi^- p \to \pi^- p$ is essentially imaginary, positive and slowly dependent on energy variable $s$ at small $t$ and that it dominates the helicity flip amplitude $H_{0+,0-}$. These requirements are in agreement with data on $\sigma_{tot}(\pi^- p \to \pi^- p)$ through optical theorem and with measurements of target polarization and recoil nucleon polarization in $\pi^- p \to \pi^- p$ from 6 to  40 GeV/c\cite{pierrard75}. The amplitude analysis of $\pi^{\pm} p \to \pi^{\pm} p$ reactions using fixed-$t$ dispersion relations and all available spin measurements from 10 - 100 GeV/c is also in agreement with these requirements\cite{pierrard78}.    

The off-shell amplitudes $T^J_\pi(m^2,t,0)$ are related to the on-shell partial wave amplitudes $f^J_\pi(m^2)$ in $\pi^-\pi^+ \to \pi^- \pi^+$ scattering\cite{martin76}

\begin{equation}
T^J_\pi(m^2,t,0)=\Bigl({q_\pi(off) \over q_\pi(on)}\Bigr)^J f^J_\pi(m^2)  
\end{equation}
\noindent
where $q_\pi(off)$ and $q_\pi(on)$ are off-shell and on-shell pion momenta with

\begin{equation}
4m^2q^2_\pi(off)=(t-(m+\mu)^2)(t-(m-\mu)^2)
\end{equation}
\noindent
The connection of pion production to $\pi\pi$ scattering through factorization (6.3) was first proposed by Goebel\cite{goebel58} and Chew and Low\cite{chew59}. The possible importance of the non-pole term was emphasized early by Gutay\cite{gutay69}. However it is generaly accepted that the non-pole term is small at small t\cite{martin76}. To the extent that the non-pole term can be neglected at small $t$, the $S$-wave partial wave amplitude in $\pi^- \pi^+ \to \pi^- \pi^+$ imparts a phase to the production amplitude $S_1$.

Partial wave unitarity in $\pi^-\pi^+ \to \pi^- \pi^+$ scattering imposes a positivity condition on the imaginary parts of partial wave amplitudes $Im f^J_\pi > 0$\cite{svec01} which implies positivity

\begin{equation}
Im T^J_\pi(m^2,t,0) > 0
\end{equation}
\noindent
Assuming that the non-pole term is small and taking into account the positivity of the factor $F^J_1$, the unitarity condition (6.5) then implies the positivity condition 

\begin{equation}
Im A^J_1(s,t,m^2) > 0
\end{equation}
\noindent
on the imaginary parts of the helicity flip production amplitudes $A^J_1$ for any $J$.

In Figure 7 we see that $Im S_1 > 0$ for the "down" solutions $S_1(+)$ for all solution combinations $(i,\overline j)$ and for both fits A and B. For the "up" solution $Im S_1(-) = - Re S_0(+)$. From Figure 7 we see that $Im S_1(-)$ is large and negative below 900 MeV for all solution combinations. We conclude that the"up" solution is clearly excluded by the unitarity in $\pi\pi$ scattering. 

Next  we note that we can write the $a_1$ exchange nonflip amplitudes $A^J_0=H^J_{0+,0+}$ in a form similar to (6.2) with $a_1$ pole term and use a factorization similar to (6.3) to connect the production amplitudes $A^J_0$ to off-shell partial wave amplitudes $T^J_{a_1}(m^2,t,1)$ in $\pi^- a_1^+ \to \pi^-\pi^+$ scattering. The positivity of $Im T^J_{a_1}(m^2,t,1)$ from unitarity in  $\pi^- a_1^+ \to \pi^-\pi^+$ then implies positivity $Im A^J_0(s,t,m^2) > 0$. This condition is in full agreement with our results for $Im S_0$ shown in Figure 7. This agreement  indicates consistency of the "down" solution and the evidence for $\sigma(770)$ resonance with unitarity in both $\pi^- \pi^+ \to \pi^-\pi^+$ and $\pi^- a_1^+ \to \pi^-\pi^+$ scattering. 

A detailed look at $Im S_1$ in Figure 7 shows small negative values at $\sim$ 980 MeV in the Fit A in two solution combinations $(1, \overline 2)$ and  $(2, \overline 2)$. $Im S_0$ shows similar negative values in Fit A in all solution combinations except  $(2, \overline 1)$. This may indicate a small negative contribution from the non-pole terms. However we cannot exclude a possibility of a small violation of unitarity at this mass in the Fit A in these solution combinations. In such a case the unitarity would favour the Fit B.    

The production amplitudes $H^J_{0 \lambda_n, 0 \lambda_p}$ can be considered as helicity amplitudes in a two-body scattering $a+b \to c(J,m) +d$ where the particle c has spin $J$ and mass $m$. A more rigorous treatment replaces the elementary particle exchange with a leading Reggeon exchange. The exchanged particle propagators are replaced by signature factors with a Regge energy dependence, and the amplitudes $T^J_{\pi}(m,t,0)$ and $T^J_{a_1}(m,t,1)$ are replaced by pion - Reggeon scattering partial wave amplitudes $T^J_{\pi}(m,t,\alpha_\pi)$ and $T^J_{a_1}(m,t,\alpha_{a_1})$ with Regge trajectories $\alpha_\pi(t)$ and $\alpha_{a_1}(t)$\cite{perl74,irving77}. This formalism has the advantage that it can be applied also to unnatural and natural exchange amplitudes which are combinations of helicity amplitudes with
 $\lambda \ne 0$. To the extent that the non-pole terms can be neglected at small $t$, the analyticity of production amplitudes can be related to the analyticity of the partial wave amplitudes in pion - Reggeon scattering.

Finally we comment on the $S$-wave $\pi \pi$ phase shift $\delta^0_0$. In Ref.~\cite{svec02} we show that the preferred solution combination is $(1, \overline 1)$. We thus focus on the results in Figure 7a. Neglecting isospin $I=2$ contribution to the $S$-wave partial wave $f^0_{\pi}$ and assuming that the inelasticity $\eta^0_0=1$ below 900 MeV, we get $f^0_{\pi} = {1 \over 3}f^0_0$ where the isospin $I=0$ amplitude

\begin{equation}
f^0_0 = {1 \over q} \sin \delta^0_0 e^{i \delta^0_0} = {1 \over q}(\sin \delta^0_0 \cos \delta^0_0 + i \sin^2 \delta^0_0)
\end{equation}
\noindent
Neglecting further the non-pole term $M^0_1$ in (6.2), we observe from the Figure 7a that $\delta^0_0 > 90^{\circ}$ since $Re S_1 < 0$. In contrast, the CERN-Munich phase shift $\delta^0_0$ is a rising function of $m$ with $\delta^0_0$ starting well below $90^{\circ}$ in this mass range\cite{hyams73,esta73,esta74,au87,morgan93}. Our helicity amplitudes are determined up to an overall phase factor $e^{i \phi}$. With $\phi = - {\pi \over 2}$ we get new amplitudes $\tilde S_n = -i S_n$, $n=1,2$.  In this case $Re \tilde S_1 = Im S_1 > 0$ and $Im \tilde S_1 = -Re S_1 > 0$. As the result, the unitarity in $\pi^- \pi^+ \to \pi^- \pi^+$ is still satisfied and the phase shift $\delta^0_0$ is shifted by $90^{\circ}$ so that $\delta^0_0 < 90^{\circ}$. However now $\ Im \tilde S_0 = - Re S_0 < 0$ below 800 MeV and the unitarity in $\pi^- a^+_1 \to \pi^- \pi^+$ is thus violated. All CERN-Munich determinations of $\pi \pi$ phase shifts work with amplitude $\tilde S_1$ in the $s$-channel\cite{martin76}. The issue of unitarity in 
$\pi^- a^+_1 \to \pi^- \pi^+$ did not arise in these determinations of $\pi \pi$ phase shifts since all these analyses assume $\tilde S_0 \equiv 0$.

\section{Comparison with $\pi^0\pi^0$ mass spectra}

A comparison of  $S$-wave intensities $I_S$ from measurements of  $\pi^- p \to \pi^- \pi^+ n$ and $\pi^+n \to \pi^+ \pi^- p$ on polarized targets at four different energies in three different experiments shows a clear evidence for $\sigma(770)$ in all data - see Figure 2 and 3  above for $\pi^- p \to \pi^- \pi^+ n$ and Figures 9 and 10 of Ref.~\cite{svec97c} for $\pi^+ n \to \pi^+ \pi^- p$. In contrast,  recent measurements of  $\pi^- p \to \pi^0 \pi^0 n$ by E852 Collaboration at BNL at 18.3 GeV/c\cite{gunter01} indicate a suppression of $\sigma(770)$ in $S$-wave intensity in $\pi^0 \pi^0$ production and a possible presence of $\sigma(500)$ resonance. 

All these data come from very high quality measurements and the differences between them cannot be used to exclude one or the other. We suggest that these results are all real. The unexpected  differences in the charged and neutral pion production data may be due to differences in hadron production mechanism to which resonances are closely connected and about which they provide relevant information. 

This view is supported by other experimental facts. The $S$-wave intensity in $\pi^- p \to \pi^0 \pi^0 n$ at 18.3 GeV shows significant variations in momentum transfer $-t$\cite{gunter01}. Similar strong $t$-dependence resembling an oscillatory behaviour was observed in the $P$-wave amplitudes in $\pi^+ n \to \pi^+ \pi^- p$ and $K^+ n \to K^+ \pi^- p$\cite{svec90}. 
The data show also important differences in the $D$-wave contributions to $\pi^- p \to \pi^- \pi^+ n$ and $\pi^- p \to \pi^0 \pi^0 n$ below 1000 MeV at small $t$. While for dipion helicity $\lambda =0$ the intensity $I_{D^0}$ is near zero in both reactions, this is not the case for $\lambda \ne 0$. The corresponding natural and unnatural exchange $D$-wave intensities $I_{D^+}$ and $I_{D^-}$ are both near zero in $\pi^- p \to \pi^- \pi^+ n$ but make a significant nonzero contribution in $\pi^- p \to \pi^0 \pi^0 n$\cite{gunter01}.    

Additive Quark Model (AQM) provides a model of hadron scattering in nonperturbative region. AQM relates spin amplitudes in $K^+ n \to K^{*0} p$ and $p p \to \Delta^{++}n$ which lead to relations between spin observables in $K^+ n \to K^+ \pi^- p$ and $p p \to p \pi^+ n$. These relations are in a remarkable agreement with experiments\cite{svec89} and demonstrate in another way the close relationship between resonances and the production mechanism . 

Integrated $\pi^0\pi^0$ mass spectra were also measured by Crystal Barrel Collaboration in $\overline p p \to 3\pi^0$\cite{amsler95} and in $\overline p p \to 5\pi^0$\cite{abele96}, and by GAMS Collaboration in central production $pp \to p_f\pi^0\pi^0 p_s$\cite{alde97}. The spectra do not show narrow peaks near 750 MeV to indicate a presence of $\sigma(770)$. However the $\pi^0 \pi^0$ spectra show marked differences in all these measurements with broad peaks at 800 MeV, 600 MeV and 500 MeV, respectively. These differences also indicate that production mechanism plays a significant role in the formation of the $\pi^0\pi^0$ spectrum and challenge the naive expectation that all dipion mass spectra should show the same structure.  

Another point to consider is that scalars are probes of QCD vacuum\cite{bijmens00}. The QCD vacuum participates in hadron scattering at low as well as at high energies\cite{shifman92,kharzeev01} and the differences in production of scalar mesons in production and decay processes may reveal new information about the structure of QCD vacuum.

\section{Summary.}
 
We have performed a model independent amplitude analysis of reaction $\pi^- p\to\pi^-\pi^+ n$ at 17.2 GeV/c extending the range of dipion mass from 600--900 MeV of the earlier analysis\cite{svec96,svec97c} to 580--1080 MeV in the present analysis. We have supplemented the CERN data by the assumption of analyticity in dipion mass of the production amplitudes. This allows us to parametrize the $S$-wave transversity amplitudes $S$ and $\overline S$ in terms of Breit-Wigner amplitudes for $\sigma(770)$ and $f_0(980)$ and a coherent background and to determine these amplitudes from simultaneous fits to the moduli $|S|^2\Sigma$ and $|\overline S|^2\Sigma$. The assumed fixed $s$ and $t$ dispersion relations impart phases to production amplitudes. The phases of the fitted amplitudes are thus the true phases of the transversity amplitudes. This allows us to separate and determine the helicity nonflip and helicity flip amplitudes $S_0$ and $S_1$ corresponding to $a_1$ and $\pi$ exchange, respectively.

The sign ambiguity in the fitted amplitude $\overline S$ results in two solutions, "up" and "down", for the helicity amplitudes. The "up" solution is excluded by unitarity in $\pi \pi$ scattering. The "down" solution - and thus the evidence for $\sigma(770)$ resonance - is in agreement with unitarity in both $\pi^- \pi^+ \to \pi^- \pi^+$ and $\pi^- a^+_1 \to \pi^- \pi ^+$ scattering.  

In the "down" solution the $\sigma(770)$ is suppressed in the $\pi$ exchange amplitude $S_1$ and thus also in $\pi^- \pi^+ \to \pi^- \pi^+$ scattering while it dominates the $a_1$ exchange amplitude $S_0$.  However we find that the narrow $\sigma(770)$ manifests itself as a broad resonance at $\sim$ 720 MeV in the $\pi$ exchange amplitude $S_1$ in two solution combinations $(1, \overline 1)$ and $(2, \overline 1)$. In Ref.~\cite{svec02} we show how this dual manifestation of $\sigma(770)$ is connected to breaking of scale and chiral symmetry in QCD. The symmetries select the solution $(1, \overline 1)$ for which $m_\sigma = 769 \pm 13$ MeV and $\Gamma_\sigma = 154 \pm 22$ MeV.

The CERN measurements of pion production on polarized targets have opened a new approach to experimental hadron spectroscopy by making possible the study of resonance production on the level of spin amplitudes. The experiments also show the close connection between resonances and the production mechanism and may reveal new information about the structure of QCD vacuum. Our results emphasize the need for a dedicated and systematic study of various production processes on the level of spin amplitudes measured in experiments with polarized targets. Such "amplitude spectroscopy" will be feasible at high intensity hadron facilities\cite{svec88,svec93}. The first high intensity hadron facility will be the Japan Hadron Facility at KEK. It will become operational in 2007 and the formation of its physics program is now in progress\cite{jhf}.

\acknowledgements
I wish to thank K.~Rybicki for providing me with the numerical tables of  CERN data in 
Ref.~\cite{chabaud83,kaminski97} which enabled this study at higher masses. I am also grateful to J.~Bystricky who explained to me how to use the program FUMILI for multifunction optimization needed for simultaneous fits.

\begin{figure}
\caption{Mass dependence of physical solutions for unnormalized moduli squared of $S$-wave and $P$-wave nucleon transversity amplitudes and cosines of their relative phases in reaction $\pi^- p \to \pi^- \pi^+ n$ at 17.2 GeV/c and momentum transfers $-t=0.005-0.2$ (GeV/c)$^2$. The results are in the $t$-channel dipion helicity frame.}\label{fig1}
\end{figure}

\begin{figure}
\caption{ Four solutions for the $S$-wave partial-wave intensity $I_S$ in the reaction $\pi^- p \to \pi^-\pi^+ n$ at 17.2 GeV/c and $-t=0.005-0.20$ (GeV/c)$^2$.}\label{fig 2}
\end{figure}
 
\begin{figure}
\caption{Four solutions for the $S$-wave partial-wave intensity $I_S$ in the reaction $\pi^- p \to \pi^-\pi^+ n$ at 1.78 GeV/c and $-t=0.005-0.20$ (GeV/c)$^2$. Data from Ref.~[26].}\label{fig 3}
\end{figure}

\begin{figure}
\caption{Consistency of measured amplitudes with the normalization condition (2.1) and phase conditions (2.5).}\label{fig 4}
\end{figure}
 
\begin{figure}
\caption{$S$-wave transversity amplitudes $|\overline S|^2\Sigma$ and $|S|^2\Sigma$ measured in $\pi^- p_\uparrow \to \pi^- \pi^+ n$ at 17.2 GeV/c and $-t = 0.005 - 0.20
$ (GeV/c)$^2$ using the  $\chi^2$ minimization method in amplitude analysis of Data Set 3. Data from Ref.~[19]. }\label{fig 5}
\end{figure}

\begin{figure}
\caption{$S$-wave transversity amplitudes $|\overline S|^2\Sigma$ and $|S|^2\Sigma$ measured in $\pi^- p_\uparrow \to \pi^- \pi^+ n$ at 17.2 GeV/c and $-t = 0.005 - 0.20
$ (GeV/c)$^2$ using the Monte Carlo method in amplitude analysis of Data Set 3 (this paper). The curves show the best simultaneous fits to the measured amplitudes $|\overline S|^2\Sigma$ and $|S|^2\Sigma$ in the four solution combinations $(1, \overline 1)$, $(2, \overline 1)$, $(1, \overline 2)$ and $(2, \overline 2)$ using the parametrization (4.9). The fits A and B yield virtually identical curves.}\label{fig 6}
\end{figure}
 
\begin{figure}
\caption{Helicity nonflip amplitude $S_0$ and helicity flip amplitude $S_1$ for Fits A and B in the four solution combinations $(1, \overline 1)$, $(2, \overline 1)$, $(1, \overline 2)$ and $(2, \overline 2)$. The Fit $A_m$ corresponds to the total absence of $\sigma(770)$ in the helicity flip amplitude $S_1$.}\label{fig 7}
\end{figure}

\begin{table}
\caption{Mass and width of $\sigma(770)$ and $f_0(980)$ from simultaneous fits to $|\overline S|^2\Sigma$ and $|S|^2\Sigma$. The combination of solutions of $|\overline S|^2\Sigma$ and $|S|^2\Sigma$ is given in the parentheses on the left with the notation $\overline 1$ and $\overline 2$ for solutions of amplitude $|\overline S|^2\Sigma$ and 1 and 2 for solutions of amplitude $|S|^2\Sigma$. Fits A and B give the same results.}\label{table1}
\begin{tabular}{cccccc}
{Fit} &{$m_\sigma$} &{$\Gamma_\sigma$} &{$m_f$} &{$\Gamma_f$} &{$\chi^2$/dof}\\
 &{(MeV)} &{(MeV)} & {(MeV)} & {(MeV)}\\
\tableline
$(1, \overline 1)$ & 769 $\pm$ 13 & 154 $\pm$ 22 & 979 $\pm$ 12 & 60 $\pm$ 23 
&\dec 0.5395 \\
$(2, \overline 1)$ & 774 $\pm$ 12 & 121 $\pm$ 23 & 956 $\pm$ 16 & 74 $\pm$ 32 
&\dec 0.3423 \\
$(1, \overline 2)$ & 787 $\pm$ 19 & 165 $\pm$ 45 & 982 $\pm$  6 & 27 $\pm$ 26 
&\dec 0.5027 \\
$(2, \overline 2)$ & 780 $\pm$ 18 & 126 $\pm$ 40 & 975 $\pm$ 14 & 44 $\pm$ 20 
&\dec 0.3836 \\
\end{tabular}
\end{table}

\begin{table}
\caption{Normalization, background and $f_0(980)$ coupling parameters for amplitudes $|\overline S|^2\Sigma$ and $|S|^2\Sigma$ from simultaneous fits: Results for Fit  A. The combinations of solutions of $|\overline S|^2\Sigma$ and $|S|^2\Sigma$ as in Table I.}\label{table3}
\begin{tabular}{ccccc}
Fit & $(1, \overline 1)$ & $(2, \overline 1)$ & $(1, \overline 2)$ & $(2, \overline 2)$ \\
$\overline N_S$ & 3.652 & 3.239 & 1.564 & 1.478\\
$N_S$ & 1.966 & 1.560 &2.712 & 1.826\\
$\overline B_1$ & 0.522 & 0.693 & 0.997 & 1.128\\
$\overline B_2$ & 0.260 & 0.183 & 0.881 & 0.893\\
$B_1$ & -0.319 & -0.436 & -0.118 & -0.370\\
$B_2$ & 0.332 & 0.656 & 0.211 & 0.633\\
$\overline C_1$ & -0.399 & 0.033 & -0.627 & -0.782\\
$\overline C_2$ & 0.499 & 0.929 & 1.961 & 2.392\\
$C_1$ & -0.796 & -0.439 & -0.838 & -0.533\\
$C_2$ & 0.065 & 0.622 & 0.161 & 0.466\\
\end{tabular}
\end{table}

\begin{table}
\caption{Normalization, background and $f_0(980)$ coupling parameters for amplitudes $|\overline S|^2\Sigma$ and $|S|^2\Sigma$ from simultaneous fits: Results for Fit B. The combinations of solutions of $|\overline S|^2\Sigma$ and $|S|^2\Sigma$ as in Table I.}\label{table2}
\begin{tabular}{ccccc}
Fit & $(1, \overline 1)$ & $(2, \overline 1)$ & $(1, \overline 2)$ & $(2, \overline 2)$ \\
$\overline N_S$ & 3.484 & 2.985 & 1.524 & 1.382\\
$N_S$ & 1.966 & 1.556 &2.705 & 1.829\\
$\overline B_1$ & 0.518 & 0.693 & 0.987 & 1.068\\
$\overline B_2$ & 0.297 & 0.280 & 0.919 & 1.035\\
$B_1$ & -0.319 & -0.437 & -0.119 & -0.367\\
$B_2$ & 0.332 & 0.657 & 0.213 & 0.632\\
$\overline C_1$ & -0.190 & 0.244 & -0.113 & 0.359\\
$\overline C_2$ & 0.386 & 0.595 & 1.644 & 1.424\\
$C_1$ & -0.796 & -0.440 & -0.832 & -0.532\\
$C_2$ & 0.064 & 0.624 & 0.162 & 0.468\\
\end{tabular}
\end{table}

\begin{table}
\caption{Couplings of $\sigma(770)$ in helicity non-flip and flip amplitudes $S_0$ and $S_1$ in the "down" solution for Fits A and B. Solution combinations as in Table I.}\label{table4}
\begin{tabular}{cccccc}
{Fit} &{$R_{\sigma 0}$} &{$R_{\sigma 0}$} &{$R_{\sigma 1}$} &{$R_{\sigma 1}$} &{$\chi^2$/dof}\\
 &{Fit A} &{Fit B} & {Fit A} & {Fit B}\\
\tableline
$(1, \overline 1)$ &\dec 2.343 &\dec 2.312 &\dec -0.360 &\dec -0.329 &\dec 0.5395 \\
$(2, \overline 1)$ &\dec 2.156 &\dec 2.104 &\dec -0.390 &\dec -0.340 &\dec 0.3423 \\
$(1, \overline 2)$ &\dec 2.049 &\dec 2.037 &\dec  0.280 &\dec  0.290  &\dec 0.5027 \\
$(2, \overline 2)$ &\dec1.815 &\dec1.788 &\dec  0.096 &\dec  0.125 &\dec 0.3836 \\
\end{tabular}
\end{table}

\pagebreak
\pagestyle{empty}
\begin{center}
\begin{figure}
\centerline{\epsfysize=7.5in\epsfbox{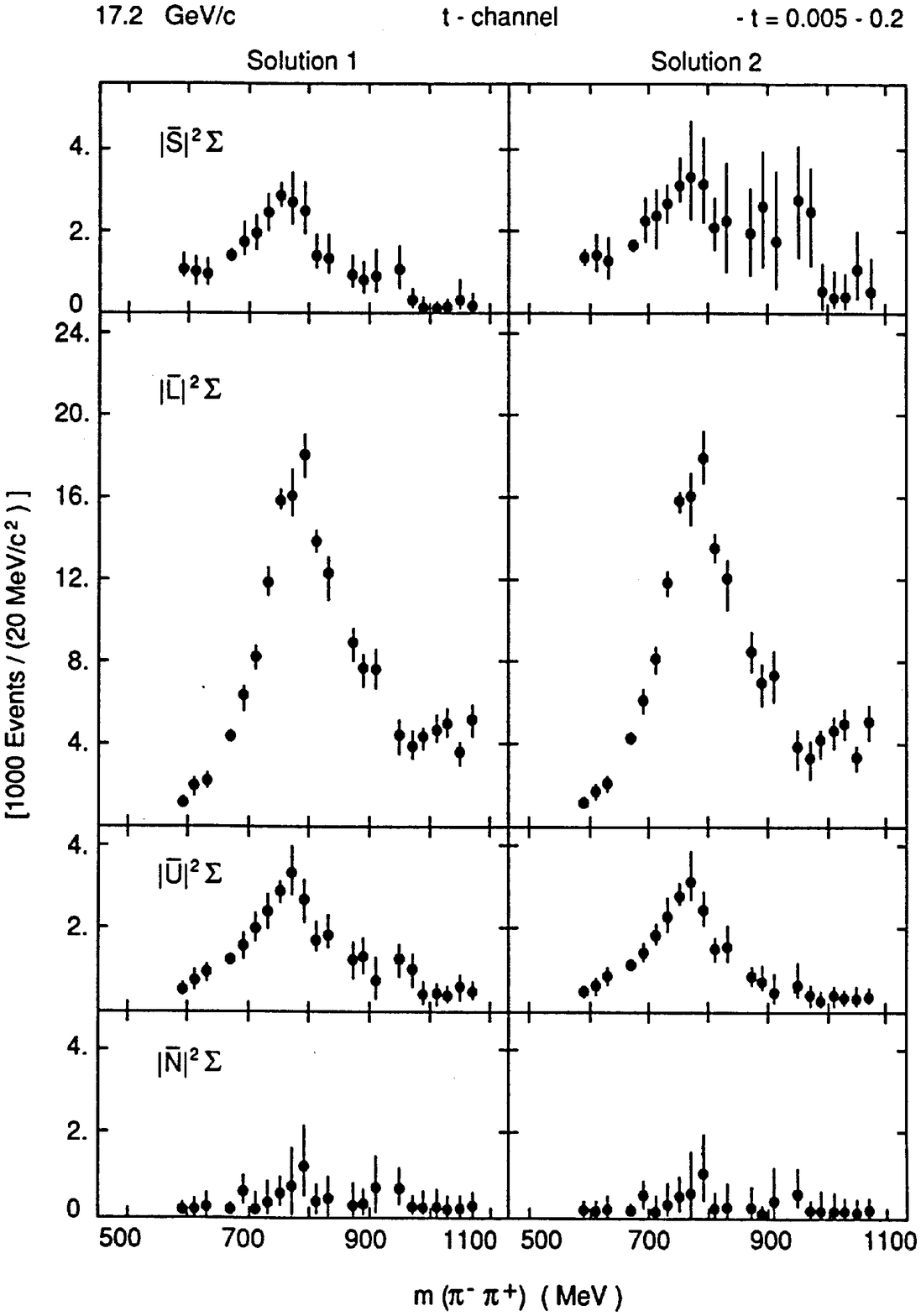}}
\bf Figure 1a
\end{figure}
\end{center}
\pagebreak
\begin{center}
\begin{figure}
\centerline{\epsfysize=7.5in\epsfbox{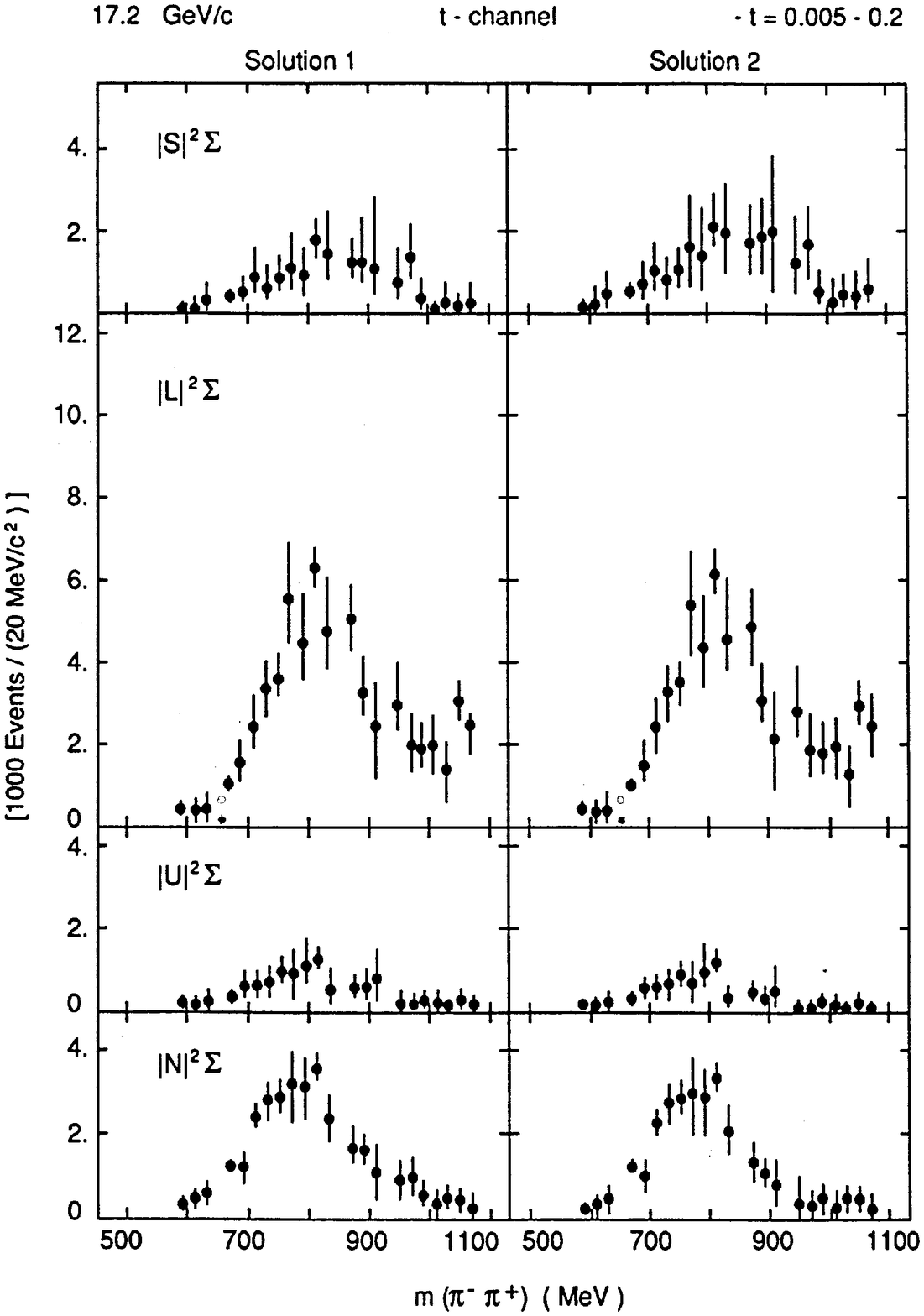}}
\bf Figure 1b
\end{figure}
\end{center}
\pagebreak
\begin{center}
\begin{figure}
\centerline{\epsfysize=7.5in\epsfbox{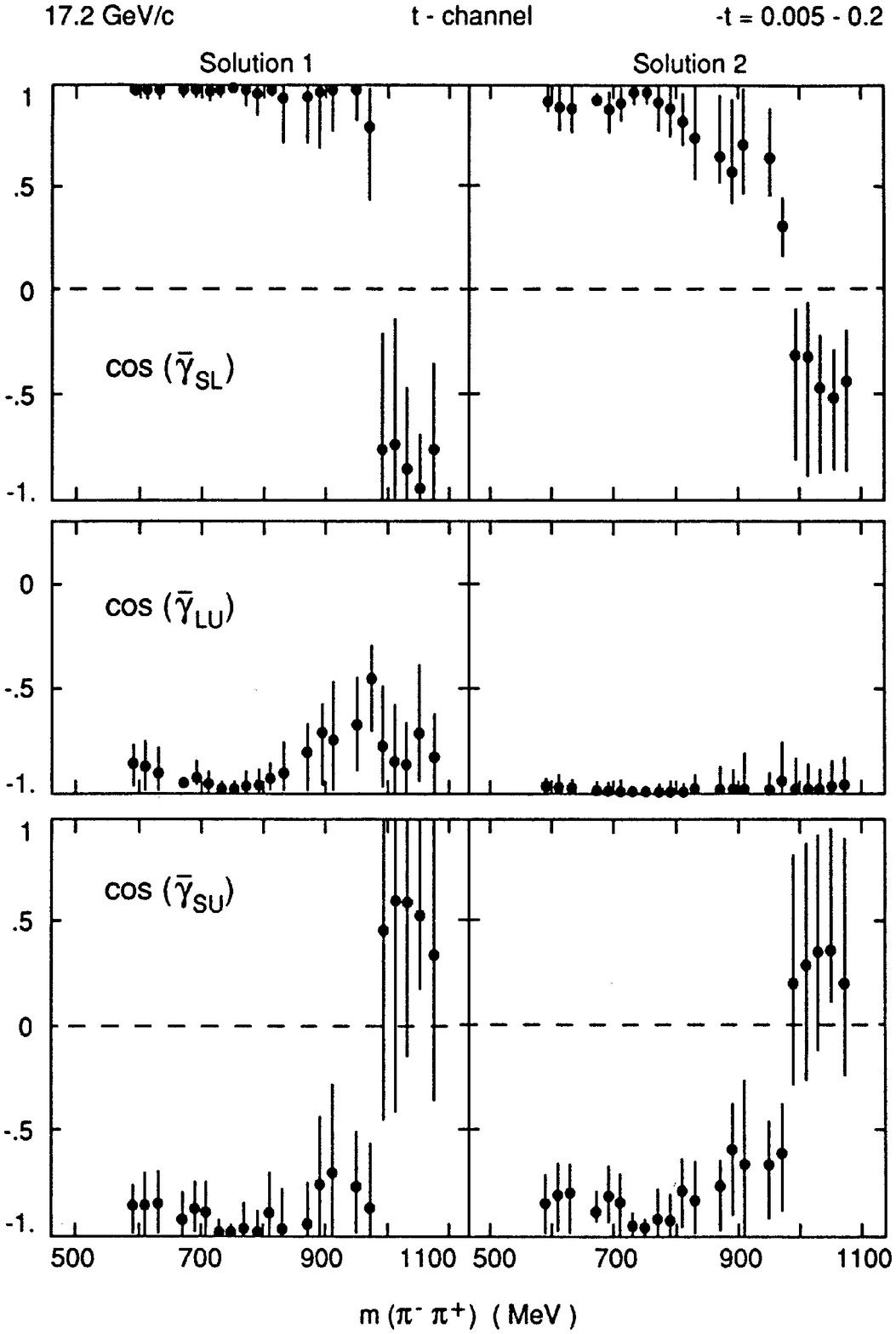}}
\bf Figure 1c
\end{figure}
\end{center}
\pagebreak
\begin{center}
\begin{figure}
\centerline{\epsfysize=7.5in\epsfbox{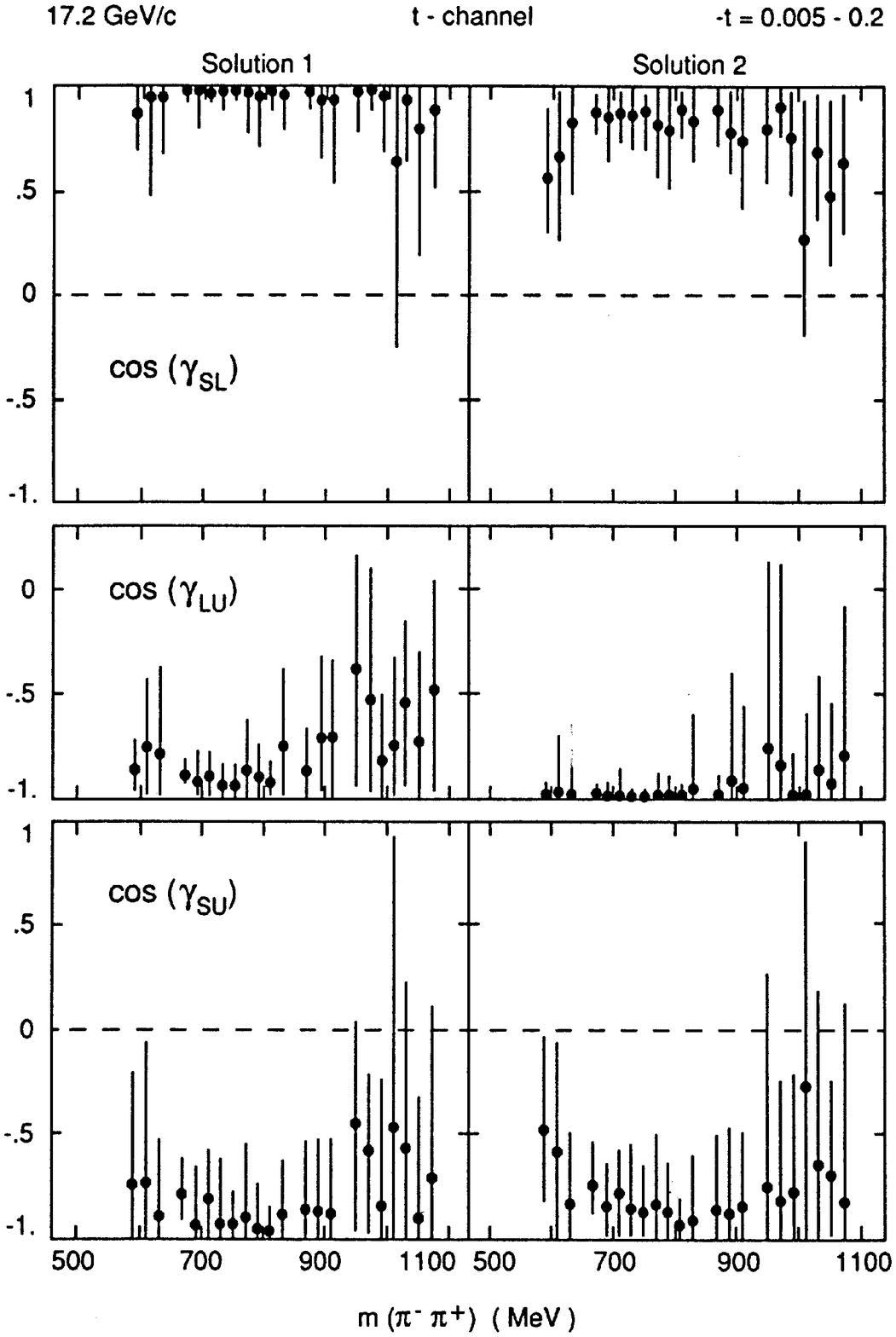}}
\bf Figure 1d
\end{figure}
\end{center}
\pagebreak
\begin{center}
\begin{figure}
\centerline{\epsfysize=7.5in\epsfbox{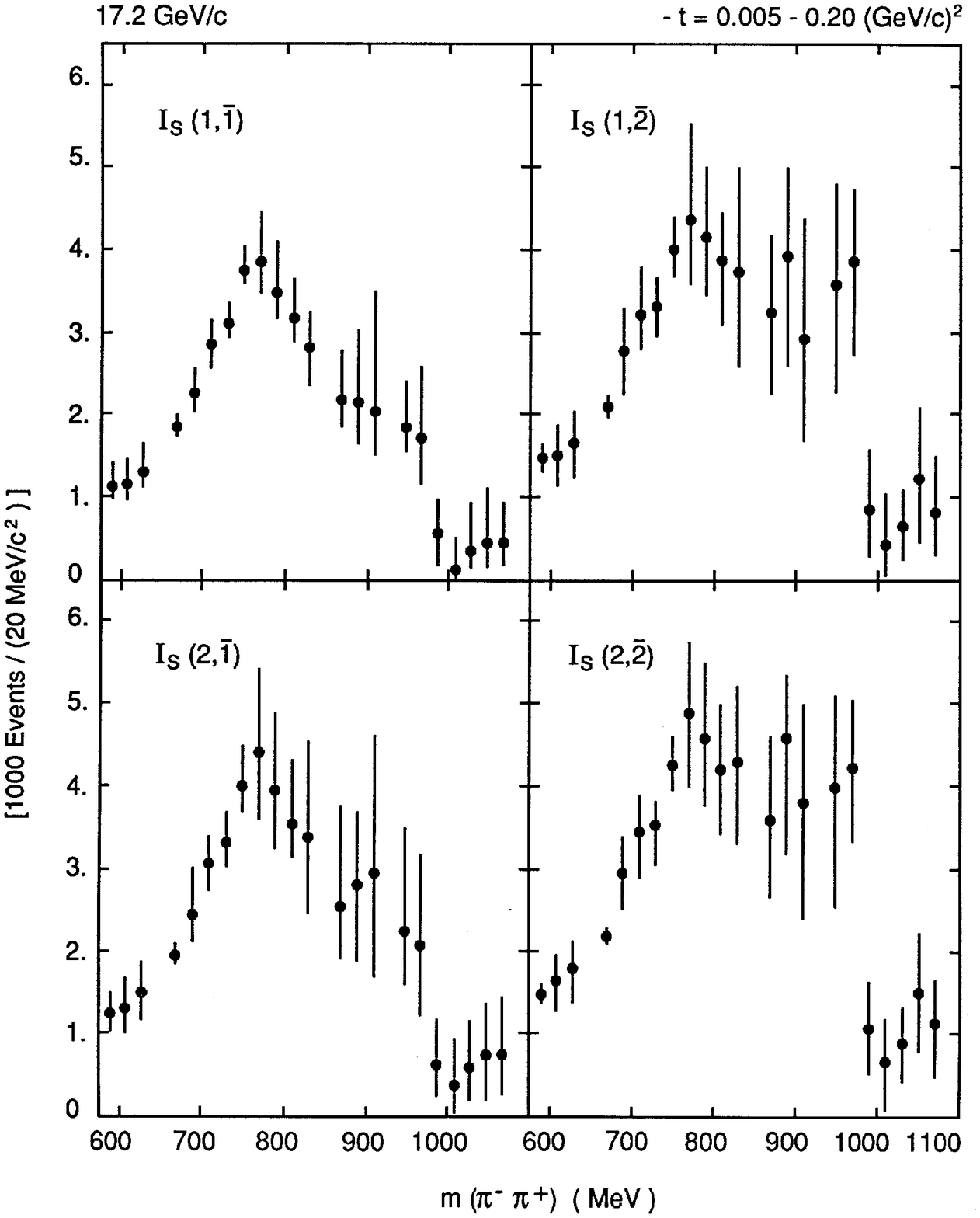}}
\bf Figure 2
\end{figure}
\end{center}
\pagebreak
\begin{center}
\begin{figure}
\centerline{\epsfysize=7.5in\epsfbox{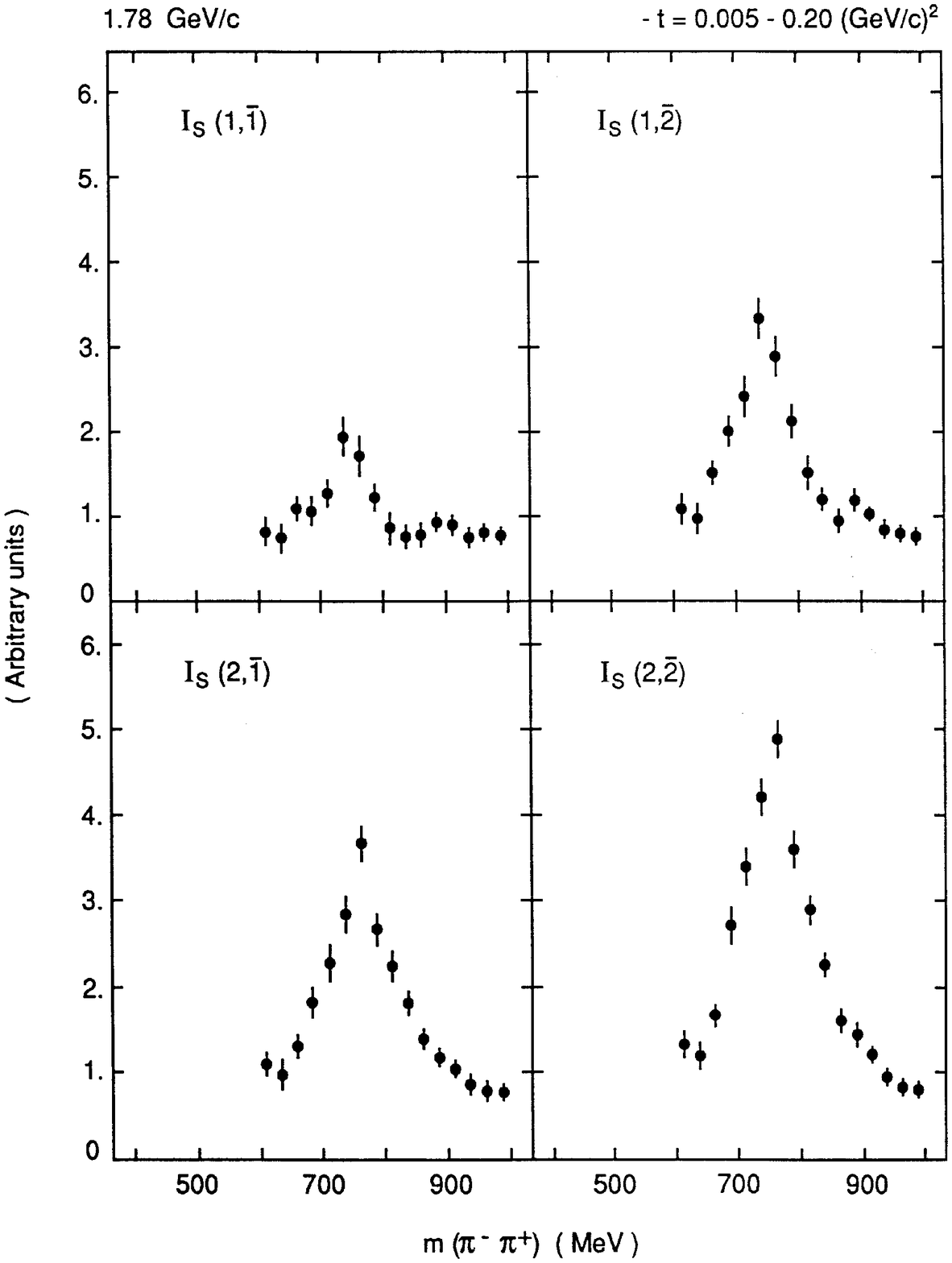}}
\bf Figure 3
\end{figure}
\end{center}
\pagebreak
\begin{center}
\begin{figure}
\centerline{\epsfysize=7.5in\epsfbox{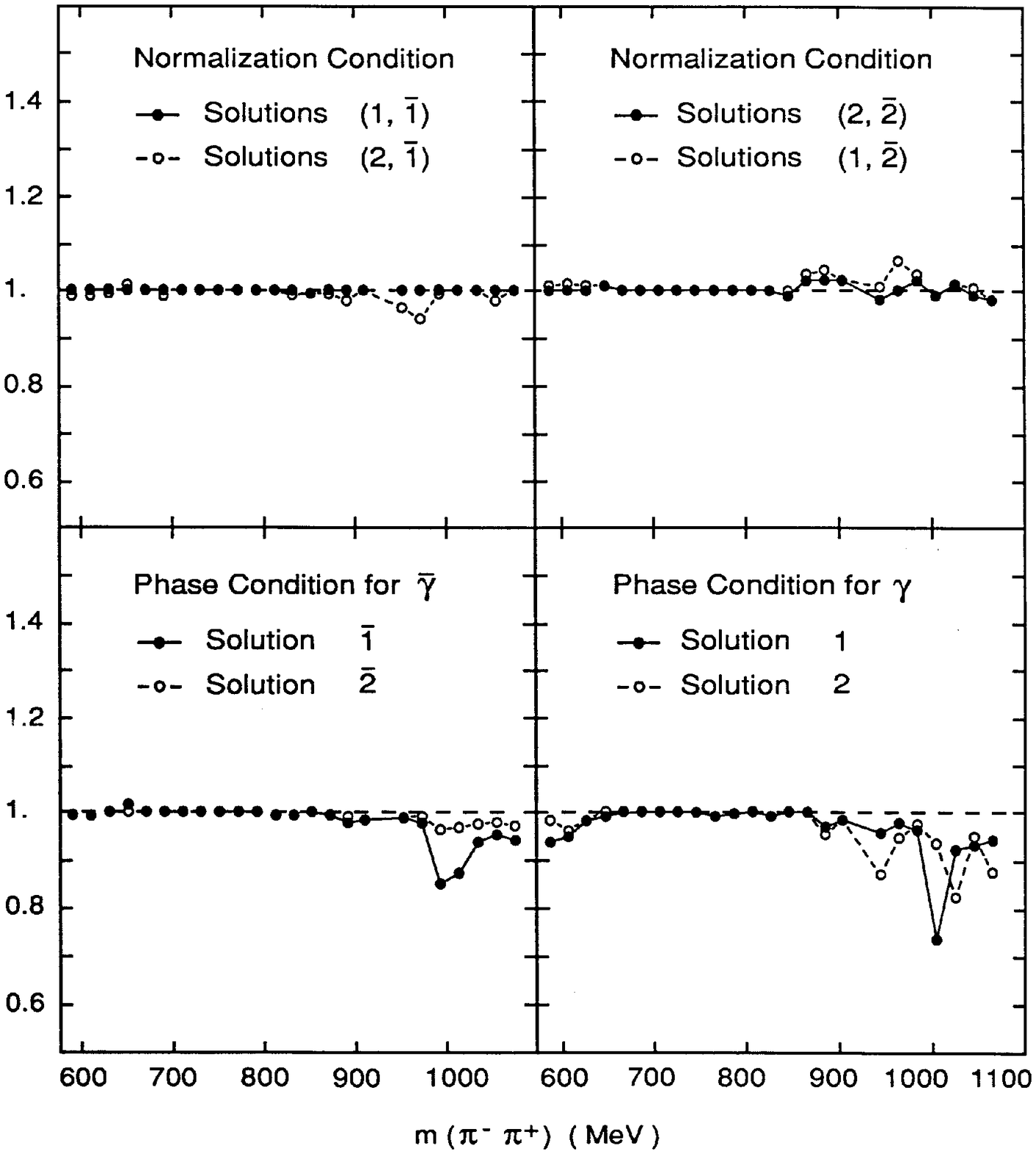}}
\bf Figure 4
\end{figure}
\end{center}
\pagebreak
\begin{center}
\begin{figure}
\centerline{\epsfysize=7.5in\epsfbox{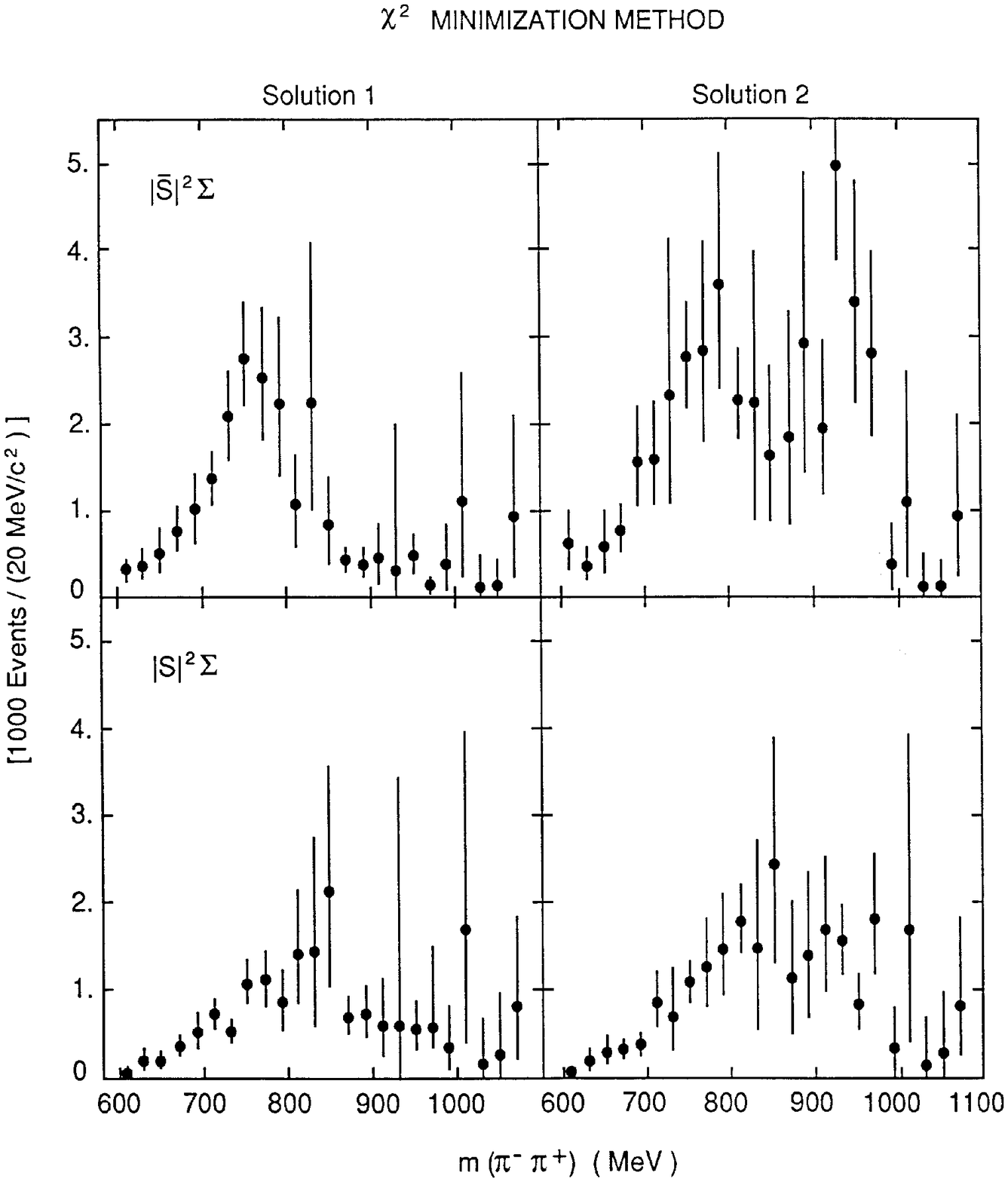}}
\bf Figure 5
\end{figure}
\end{center}
\pagebreak
\begin{center}
\begin{figure}
\centerline{\epsfysize=7.5in\epsfbox{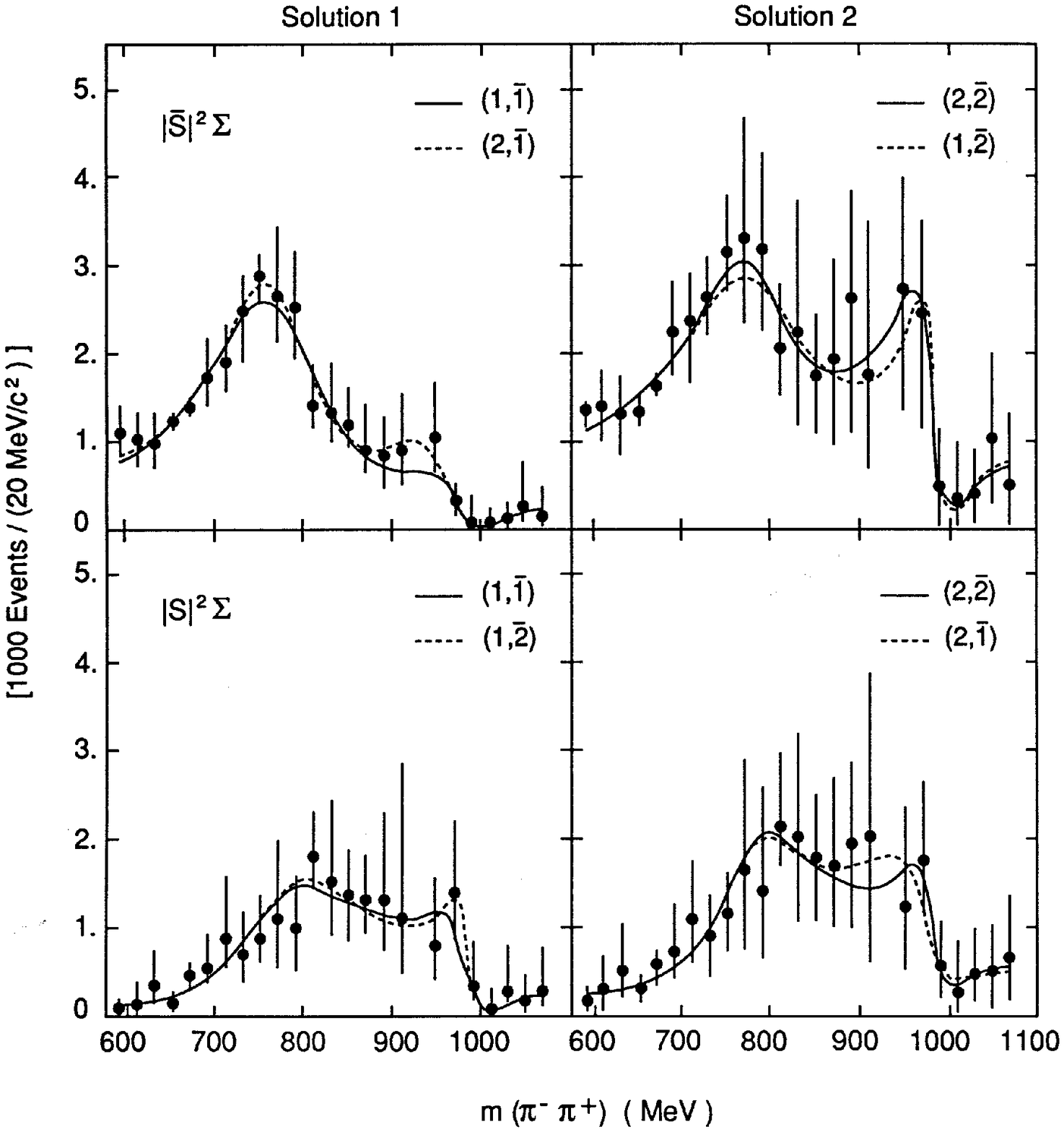}}
\bf Figure 6
\end{figure}
\end{center}
\pagebreak
\begin{center}
\begin{figure}
\centerline{\epsfysize=7.5in\epsfbox{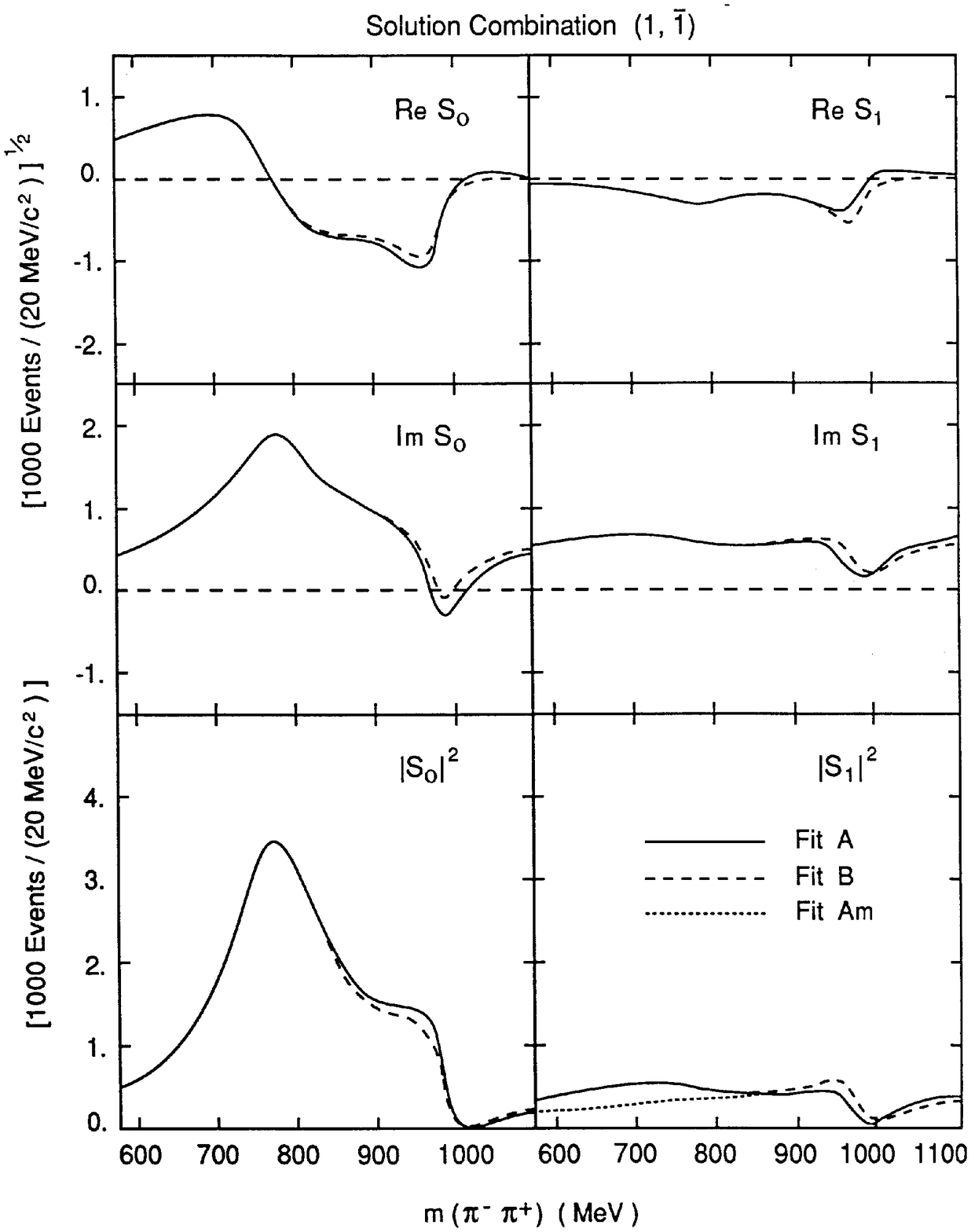}}
\bf Figure 7a
\end{figure}
\end{center}
\pagebreak
\begin{center}
\begin{figure}
\centerline{\epsfysize=7.5in\epsfbox{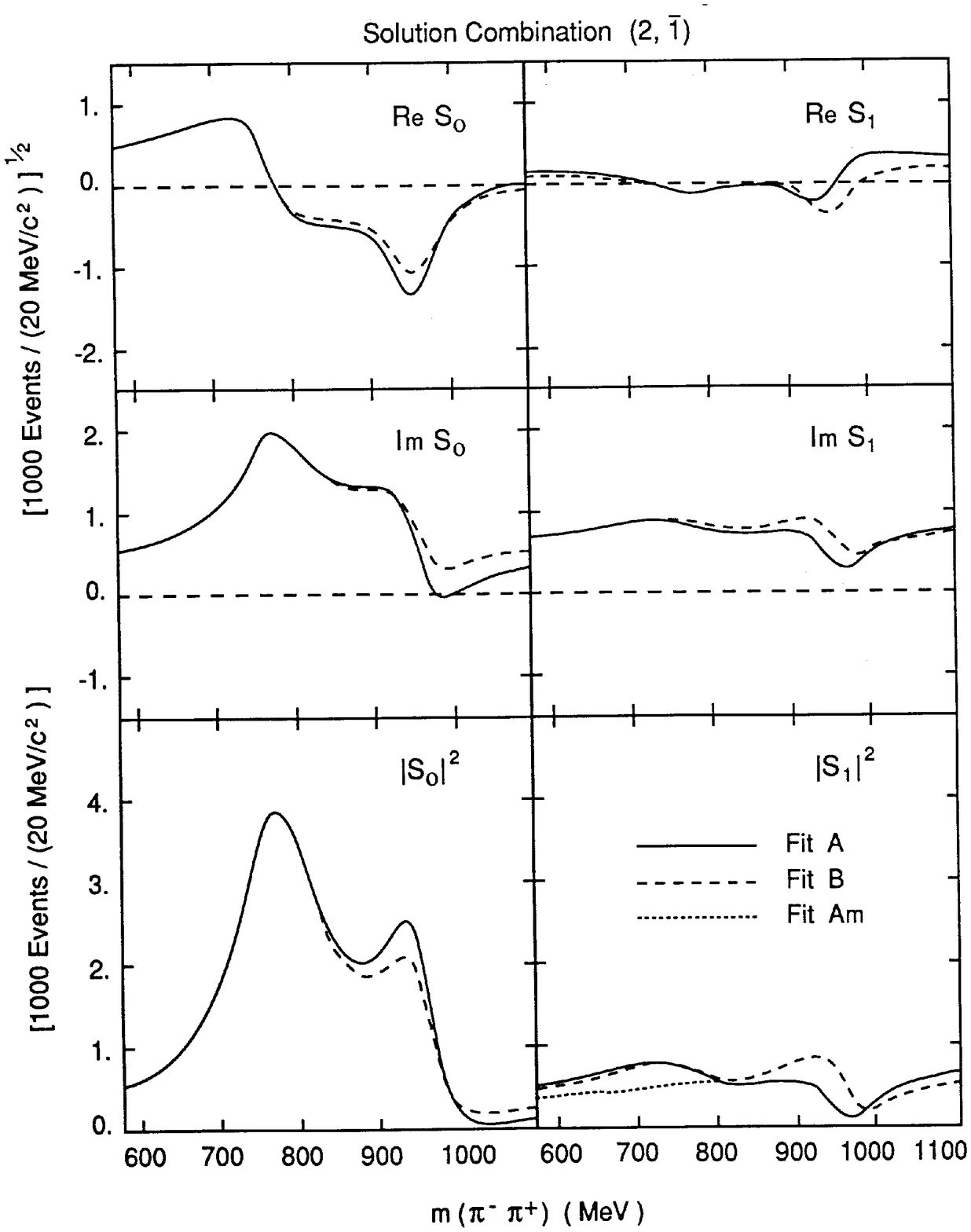}}
\bf Figure 7b
\end{figure}
\end{center}
\pagebreak
\begin{center}
\begin{figure}
\centerline{\epsfysize=7.5in\epsfbox{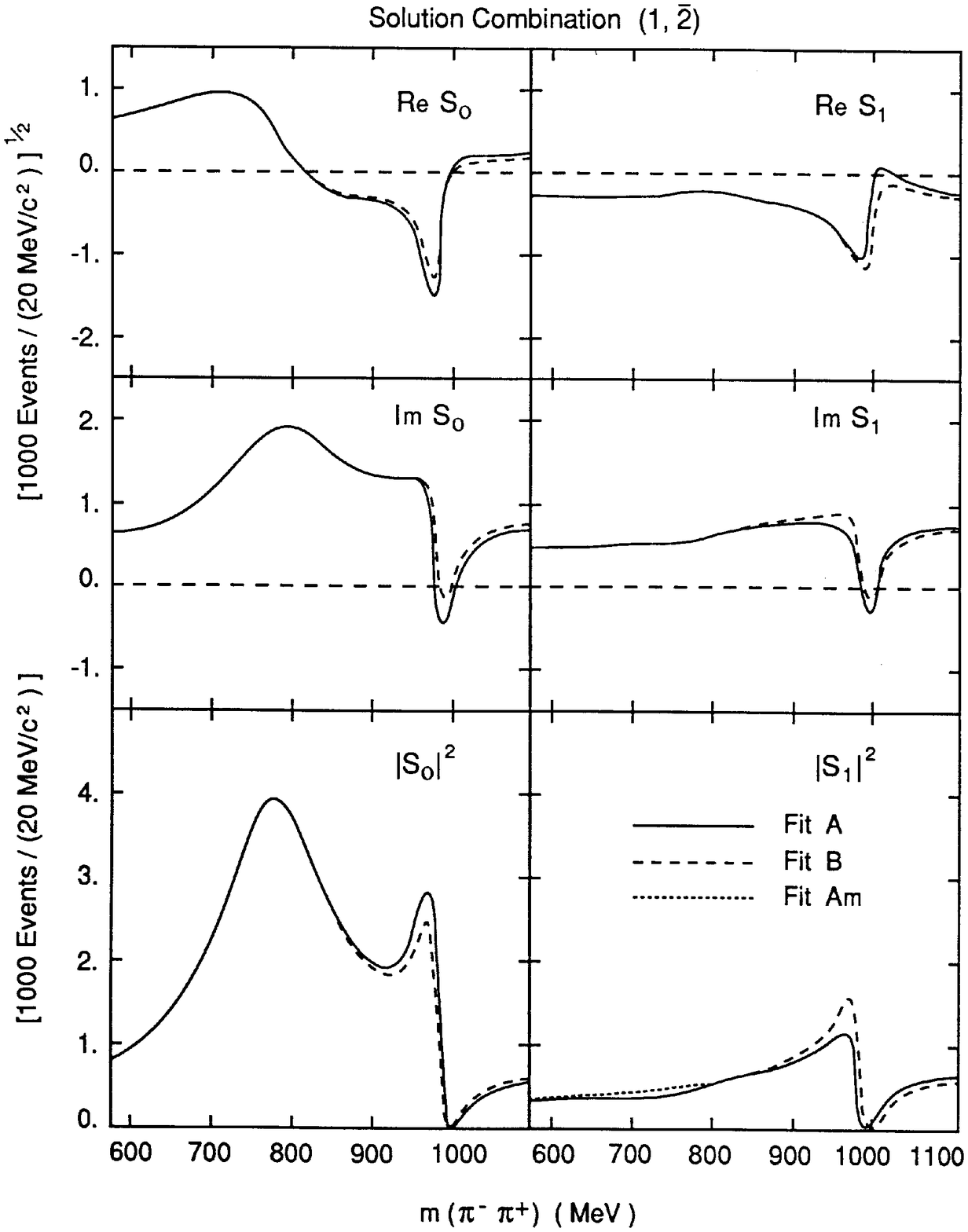}}
\bf Figure 7c
\end{figure}
\end{center}
\pagebreak
\begin{center}
\begin{figure}
\centerline{\epsfysize=7.5in\epsfbox{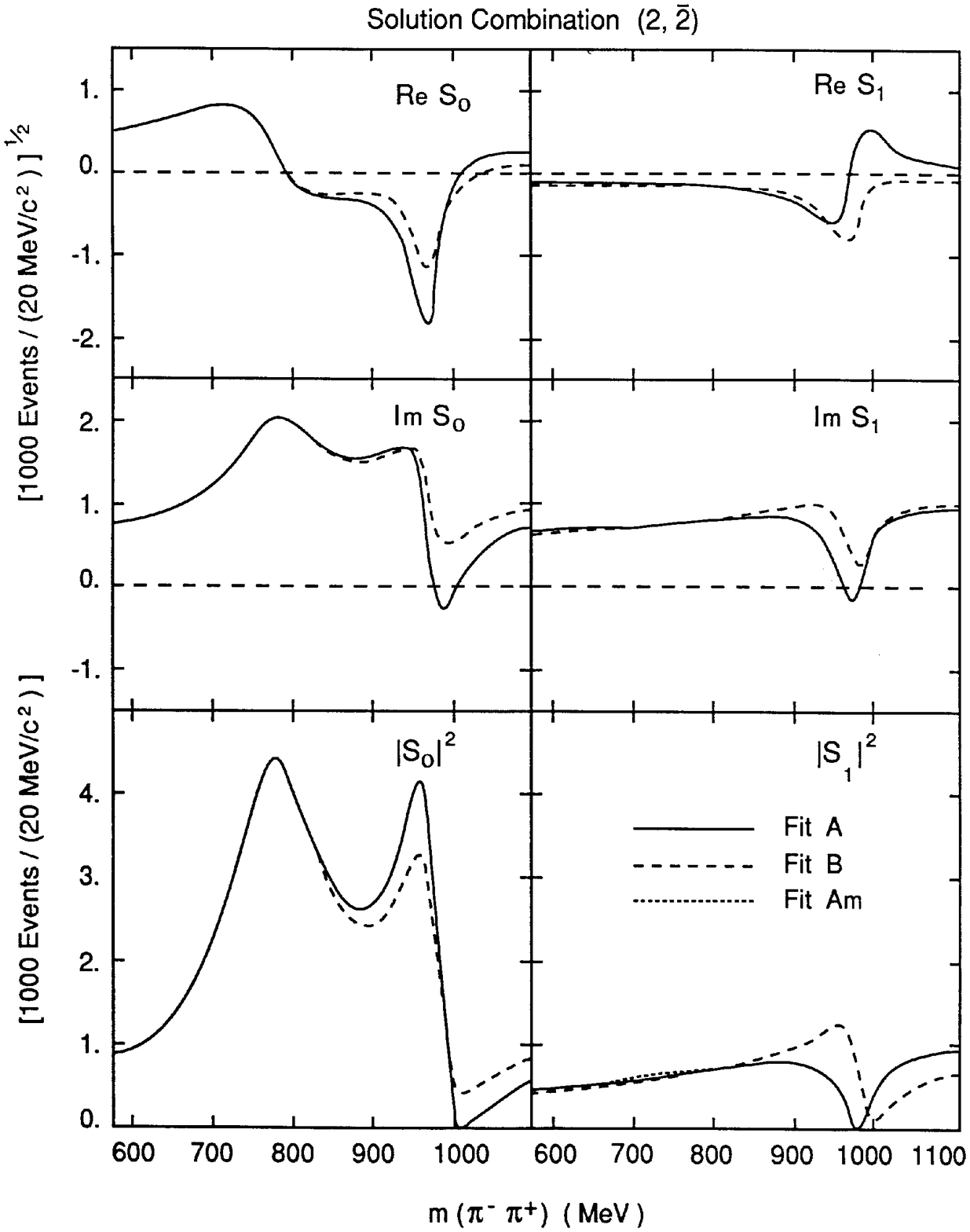}}
\bf Figure 7d
\end{figure}
\end{center}
\end{document}